\documentclass[12pt]{article}
\usepackage{amsmath}
\usepackage{longtable}
\usepackage{graphicx,psfrag,epsf}
\usepackage{enumerate}
\usepackage{natbib}
\usepackage{setspace}
\usepackage{hyperref} 
\usepackage{authblk}
\usepackage{hyperref}
\usepackage{booktabs}
\usepackage{subcaption}
\DeclareUnicodeCharacter{2212}{-}
\usepackage{url} 
\newcommand{\blind}{0}
\usepackage{amssymb}
\newcommand{\bfy}{\mathbf{y}}

\newcommand{\bfmu}{\pmb{\mu}}
\newcommand{\bfeta}{\pmb{\eta}}

\newcommand{\FMAE}{{\text{FMAE}}}

\newcommand{\FRMSE}{{\text{FRMSE}}}
\newcommand{\Normal}{\mathcal{N}}
\newcommand{\HalfNormal}{\mathrm{Half\!-\!Normal}}
\newcommand{\diag}{\operatorname{diag}}
\newcommand{\vecop}{\operatorname{vec}}
\newcommand{\LKJChol}{\mathrm{LKJ\_Corr\_Cholesky}}

\newcommand{\bfbeta}{\boldsymbol{\beta}}
\newcommand{\bfA}{\mathbf{A}}
\newcommand{\bfB}{\mathbf{B}}
\newcommand{\bfSigma}{\boldsymbol{\Sigma}}
\newcommand{\bfOmega}{\boldsymbol{\Omega}}
\newcommand{\bX}{\mathbf{X}}
\newcommand{\bY}{\mathbf{Y}}
\providecommand{\Normal}{\mathcal{N}}
\providecommand{\HalfNormal}{\mathcal{HN}}
\providecommand{\LKJChol}{\mathrm{LKJ\_Chol}}
\providecommand{\diag}{\operatorname{diag}}
\providecommand{\iid}{\;\stackrel{\text{iid}}{\sim}\;}
\providecommand{\vecop}{\mathrm{vec}}
\providecommand{\bfeta}{\boldsymbol{\eta}}
\providecommand{\bfbeta}{\boldsymbol{\beta}}
\providecommand{\bfSigma}{\boldsymbol{\Sigma}}
\providecommand{\bfOmega}{\boldsymbol{\Omega}}
\providecommand{\bfA}{\mathbf{A}}
\providecommand{\bfB}{\mathbf{B}}
\providecommand{\bX}{\mathbf{X}}
\providecommand{\bY}{\mathbf{Y}}
\providecommand{\bfz}{\mathbf{z}}
\providecommand{\BetaDist}{\operatorname{Beta}}

\newcommand{\bfX}{\pmb{X}}
\newcommand{\bfx}{\pmb{x}}
\newcommand{\bfZ}{\pmb{Z}}

\newcommand{\bftheta}{\pmb{\theta}}
\newcommand{\bfgamma}{\pmb{\gamma}}
\newcommand{\airbnb}{Airbnb }
\newcommand{\med}{med}
\newcommand{\airbnbp}{Airbnb}

\makeatletter
\newcommand*{\addFileDependency}[1]{
  \typeout{(#1)}
  \@addtofilelist{#1}
  \IfFileExists{#1}{}{\typeout{No file #1.}}
}
\makeatother

\newcommand{\alr}{{\textnormal{alr}}}
\newcommand{\clr}{{\textnormal{clr}}}

\title{A Bayesian Dirichlet  Auto-Regressive Conditional Heteroskedasticity Model for Compositional Time Series}

\setcitestyle{authoryear,open={(},close={)}}
\date{} 
\begin{document}

\def\spacingset#1{\renewcommand{\baselinestretch}%
{#1}\small\normalsize} \spacingset{1}


\if0\blind
{
  \title{\bf A Bayesian Dirichlet  Auto-Regressive Conditional Heteroskedasticity Model for Forecasting Currency Shares}
\author[1,2]{Harrison Katz}
\author[3]{Robert E. Weiss}

\affil[1]{Department of Statistics, UCLA}
\affil[2]{Forecasting, Data Science, \airbnbp}

\affil[3]{Department of Biostatistics, UCLA Fielding School of Public Health }
  \maketitle
} \fi

\if1\blind
{
  \bigskip
  \bigskip
  \bigskip
  \begin{center}
    {\LARGE\bf A Bayesian Dirichlet  Auto-Regressive Conditional Heteroskedasticity Model for Forecasting Currency Shares}
\end{center}
  \medskip
} \fi

\begin{abstract}
In marketplace finance, the daily mix of billing currencies is compositional data that drives forecasting, reporting, and treasury risk. We study Airbnb's currency-fee shares across four regions and present a Bayesian Dirichlet ARMA model with a time-varying precision component. The model keeps predictions on the simplex, captures mean dynamics on the additive log-ratio scale, and lets volatility spike during disruptions and settle as conditions normalize. We evaluate against standard Dirichlet and transformed-Gaussian alternatives using simulations with misreported observations and temporary regime shifts, then validate on held-out data. Across all settings, our approach delivers more accurate forecasts, better-calibrated intervals, and weaker residual persistence. Modeling precision as a dynamic process provides a practical, interpretable way to forecast proportions and quantify uncertainty when the noise itself moves.

\bigskip

\end{abstract}

\section{Introduction}

Airbnb operates a two-sided marketplace with millions of listings worldwide, and it accommodates transactions in over 100 currencies. While individual hosts set their own nightly rates, the platform must still \emph{forecast} these prices to predict total daily revenue. Complicating matters, travelers can pay in a currency of their choosing (often their local currency), resulting in a \emph{dynamic compositional time series} of daily fee shares across USD, EUR, CAD, and other currencies. This currency composition is crucial for accurately converting booked stays into a consolidated U.S. Dollar (USD) amount—a requirement for company-wide financial reporting, treasury operations, and broader risk management.

To forecast revenue in USD, one must know (i) the nightly prices that hosts set in their local currency, and (ii) the fraction of total bookings ultimately paid in each currency. Even if EUR comprises only a small percentage of daily transactions, that fraction can swing markedly from day to day, amplified by large booking volumes and exchange-rate shifts. Such changes directly affect net USD receipts and thus \emph{any} revenue forecast that fails to anticipate compositional shifts may lead to significant errors. Moreover, once daily compositions are known, treasury teams can plan FX conversions more efficiently, hedge against unexpected fluctuations, or analyze how macroeconomic conditions influence guest payment preferences.

Each day’s currency proportions must lie in a \emph{simplex} (they are nonnegative and sum to one), a constraint that complicates standard time-series methods. One classical approach is to use log-ratio transformations, as pioneered by \citet{aitchison1982statistical} and extended in works such as \citet{Cargnoni1997BayesianFO}, \citet{ravishanker2001compositional}, \citet{mills2010forecasting}, and \citet{snyder2017forecasting}, among others. Transformed additive log-ratio VARMA (\emph{tVARMA}) models have been fruitfully applied to compositional data \citep{brunsdon1998time, mills2009forecasting, mills2010forecasting}.

Let $Q_{t,c}$ denote gross booking amount in currency $c$ on day $t$, and let $\mathrm{FX}_{t,c\rightarrow USD}$ be the USD price of one unit of $c$. Consolidated revenue in USD can be written as
\begin{equation}
R_t^{USD} \;=\; \sum_{c=1}^{J} Q_{t,c}\,\mathrm{FX}_{t,c\rightarrow USD}
\;=\; V_t \sum_{c=1}^{J} S_{t,c}\,\mathrm{FX}_{t,c\rightarrow USD},
\quad
S_{t,c}=\frac{Q_{t,c}}{V_t},\; V_t=\sum_{c=1}^{J}Q_{t,c}.
\label{eq:usd-decomp}
\end{equation}
Thus a USD forecast requires both a scale forecast $V_t$ (total nominal volume) and a \emph{mix} forecast $S_{t,\cdot}$ (currency shares). One can pursue a \emph{bottom‑up} route (model each $Q_{t,c}$ and normalize to obtain $S_{t,\cdot}$) or a \emph{top‑down} route (model $S_{t,\cdot}$ directly on the simplex). Our focus is the latter, because the decision‑relevant object for reporting and hedging is the \emph{composition} itself: even small swings in $S_{t,\cdot}$ materially change $R_t^{USD}$ when multiplied by large $V_t$ and moving FX rates. Empirically, the share series exhibit recurrent bursts of dispersion (see Figure~\ref{training_all}), which motivates an observation‑driven volatility mechanism on the simplex.

In contrast, the Dirichlet distribution models compositional data directly on the simplex. \citet{benjamin2003generalized} proposed frequentist Generalized ARMA models that could be extended to the simplex. Contributions to Dirichlet time series include Bayesian compositional state space models by \citet{grunwald1993time}, \citet{da2011dynamic}, and \citet{da2015bayesian}; the Bayesian Dirichlet Auto-Regressive Moving Average (B-DARMA) model by \citet{KATZ20241556,forecast7030032,forecast7040062,katz2025centeredmadirichletarma}; and frequentist Dirichlet ARMA models by \citet{zheng2017dirichlet,creus2021dirichlet}.

The generalized auto-regressive conditional heteroskedasticity (GARCH) model \citep{bollerslev1986generalized} is fundamental in modeling financial time series volatility. Extensions include non-normal likelihoods to capture heavy tails and skewness \citep{nelson1991conditional, engle2001theoretical}. \citet{bauwens2006multivariate} surveyed advancements in multivariate GARCH models.

Motivated by the need to capture both shifting means and evolving volatility in compositional time series—especially under sudden disruptions such as COVID-19—we develop a new class of models called \textbf{Bayesian Dirichlet Auto-Regressive Moving Average with Dirichlet Auto-Regressive Conditional Heteroskedasticity (B-DARCH)}. Unlike existing Dirichlet ARMA methods that assume a fixed or purely deterministic precision parameter, our framework introduces a \textit{Dirichlet Auto-Regressive Conditional Heteroskedasticity (DARCH)} component. This addition endows the standard Bayesian Dirichlet ARMA model with a GARCH-like recursion on the precision $\phi_t$, enabling it to:
\begin{enumerate}
    \item Model volatility clustering by letting $\phi_t$ depend dynamically on recent innovations and prior volatility, thereby reflecting abrupt regime changes and prolonged uncertainty.
    \item Accommodate severe shocks (e.g., the onset of COVID-19) that induce persistent or recurrent surges in variance—phenomena not well captured by constant-precision or deterministic-variance models.
    \item Preserve compositional constraints via the Dirichlet likelihood, ensuring forecasted currency shares remain valid proportions over time.
    \item Leverage a full Bayesian framework to deliver predictive distributions (and credible intervals) that adapt automatically to time-varying volatility, resulting in more robust risk assessments and interval coverage.
\end{enumerate}
These innovations allow B-DARCH to simultaneously handle both systematic and dynamic changes in compositional data, providing a flexible, unified approach to \emph{dynamic heteroskedasticity on the simplex}—a capability hitherto missing in Bayesian compositional time-series models.

We apply the B-DARCH model to Airbnb's currency fee proportions across different regions, demonstrating its effectiveness in capturing complex temporal dynamics and volatility patterns in real-world compositional data. We conduct simulation studies comparing the B-DARCH model with the standard B-DARMA and Bayesian transformed VARMA (B-tVARMA) models. The B-DARCH model consistently outperforms the other models in forecast accuracy and residual diagnostics, highlighting its robustness and applicability in scenarios with heteroskedasticity, structural breaks, and regime changes. By directly modeling compositional data within the simplex and incorporating time-varying volatility, our proposed B-DARCH model offers a powerful tool for analyzing dynamic compositional time series.

The remainder of the paper is organized as follows. The next section introduces the B-DARCH model. Section~\ref{simulation} reports simulation studies comparing B-DARCH with three alternatives: (i) a B-DARMA data model; (ii) a Bayesian transformed-data normal VARMA model; and (iii) a time-varying-parameter comparator in ALR space (B\text{-}TVP\text{-}tVARMA). Section~\ref{airbnb} analyzes the \airbnb\ data. The paper concludes with a brief discussion. Additional tables and figures appear in the online Supplementary Material (Section~6).

\section{A Bayesian Dirichlet Auto-Regressive Moving Average - Dirichlet Auto-Regressive Conditional Heteroskedasticity Model}
\label{bayesiandirichlet}
\subsection{DARMA Data model}

Consider a \( J \)-component compositional time series \( \{ \mathbf{y}_t \}_{t=1}^T \) of length $T$ indexed by $t=1,\dots,T$, where each observation \( \mathbf{y}_t = (y_{t1}, \ldots, y_{tJ})^\top \) lies in the simplex

\[
\mathcal{S}^{J-1} = \left\{ \mathbf{y} \in (0,1)^J : \sum_{j=1}^J y_j = 1 \right\}.
\]
We model \( \mathbf{y}_t \) using the Dirichlet distribution with mean vector \( \boldsymbol{\mu}_t \) and precision parameter \( \phi_t > 0 \)

\begin{align}
\mathbf{y}_t \mid \boldsymbol{\mu}_t, \phi_t \sim \operatorname{Dirichlet}\left( \phi_t , \boldsymbol{\mu}_t \right),
\end{align}
where \( \boldsymbol{\mu}_t = (\mu_{t1}, \ldots, \mu_{tJ})^\top \in \mathcal{S}^{J-1} \) with probability density function (pdf)
\begin{align}
p(\mathbf{y}_t \mid \boldsymbol{\mu}_t, \phi_t) = \frac{\Gamma\left( \phi_t \right)}{\prod_{j=1}^J \Gamma\left( \phi_t \mu_{tj} \right)} \prod_{j=1}^J y_{tj}^{\phi_t \mu_{tj} - 1},
\label{dirichlet}
\end{align}
for \( \mathbf{y}_t \in \mathcal{S}^{J-1} \).

\subsubsection{Modeling the Mean using the Additive Log-Ratio Transformation}

We model $\bfmu_t$ as a function of prior observations $\bfy_1, \ldots, \bfy_{t-1}$, prior means $\bfmu_1, \ldots, \bfmu_{t-1}$ and known covariates $\bfx_t$ in a generalized linear model framework. As $\bfmu_t$ and $\bfy$ are constrained, we model $\bfmu_t$ after reducing dimension using the \textit{additive log ratio} (alr) link 
\begin{align}
\bfeta_t \equiv \alr(\bfmu_t) = \left(\log\left(\frac{\mu_{t1}}{\mu_{tj^*}}\right),\ldots,\log\left(\frac{\mu_{tJ}}{\mu_{tj^*}}\right)\right) \label{mueta}
\end{align}
where $\bfeta_t$ is the linear predictor, a $J-1$-vector taking values in $\mathbb{R}^{J-1}$, $j^*$ is a chosen reference component $1 \le j^* \le J$, and the element of $\bfeta_t$ that would correspond to the $j^*$th element $\log (\mu_{j^*}/\mu_{j^*}) = 0$ is omitted. Given $\bfeta_t$, $\bfmu_t$ is defined by the inverse of equation \eqref{mueta} where $\mu_{tj} = \exp(\eta_{tj})/(1 + \sum_{j = 1}^{J-1} \exp(\eta_{tj}))$ for $j = 1, \dots, J$, $j \ne j^*$ and for $j = j*$, $\mu_{tj*} = 1/(1 + \sum_{j = 1}^{J-1} \exp(\eta_{tj}))$.

The ALR-transformed mean vector \(\bfeta_t\) is modeled with an \emph{observation-driven VARMA-style recursion}
that uses autoregressive and moving-average terms built from past \(\alr(\bfy)\) and past linear predictors:
\begin{align}
    \mathbf{\eta}_t =  \sum_{p=1}^P \pmb{A}_p ( \text{alr}(\mathbf{y}_{t-p}) - \mathbf{X}_{t-p}\pmb{\beta} ) + \sum_{q=1}^Q \pmb{B}_q( \text{alr}(\mathbf{y}_{t-q})- \mathbf{\eta}_{t-q}) + \mathbf{X}_t \pmb{\beta},
    \label{eta_f}
\end{align}
where $P$ is the VAR lag, $Q$ is the VMA lag, and $t = m + 1, \dots, T$ where $m = \max(P,Q)$, $\pmb{A}_p$ and $\pmb{B}_q$ are $(J-1) \times (J-1)$, \(\mathbf{X}_t \in \mathbb{R}^{J-1 \times r_\beta}\) is a matrix of deterministic covariates, and \(\pmb{\beta} \in \mathbb{R}^{r_\beta}\) is a vector of regression coefficients.

\paragraph{Interpretation.}
Equation~\eqref{eta_f} defines \(\bfeta_t\) deterministically conditional on past data and past linear predictors, 
without introducing a separate innovation term for \(\bfeta_t\). 
Hence it is an observation-driven specification in the sense of GARCH-type recursions rather than a parameter-driven state-space evolution. 
A natural alternative would be to endow \(\bfeta_t\) with a stochastic transition equation (for example, a linear or nonlinear state-space model with evolution noise) 
and integrate over the latent path. 
In our Dirichlet likelihood with the DARCH precision recursion, this approach becomes computationally onerous due to the nonlinearity of 
the simplex mapping and the volatility recursion. 
The observation-driven form in~\eqref{eta_f} keeps inference tractable while still allowing rich short-run dynamics through the AR terms in \(\alr(\bfy)\) 
and the MA terms in \(\bfeta\). 
We complement this with a transformed-Gaussian comparator that does include time-variation on the mean side via time-varying coefficients 
(B\text{-}TVP\text{-}tVARMA), thereby illustrating the trade-off between state evolution and observation-driven recursions within our framework.

\subsubsection{Precision Parameter $\phi_t$}

In the B-DARMA model, precision parameter $\phi_t$ is modeled with log link as a function of an $r_\gamma$-vector of covariates $\bfz_t$, 
\begin{align}
    \phi_t = \exp(\bfz_t\bfgamma),
    \label{phi_f}
\end{align}
where $\bfgamma$ is an $r_\gamma$-vector of coefficients. If there are no covariates, $\log \phi_t = \gamma$ for all $t$. 

\subsection{DARCH Process for Precision Parameter \(\phi_t\)}
We introduce a DARCH process for the precision parameter \(\phi_t\), modeled with a log link 
\begin{align}
    \log(\phi_t) &= \sum_{l=1}^{L} \alpha_{l} (\log(\phi_{t-l}) - \mathbf{z}_{t-l}^{\prime}\boldsymbol{\gamma}) + \sum_{k=1}^{K} \tau_{k} ( \alr(\mathbf{y}_{t-k}) - \boldsymbol{\eta}_{t-k})^{\prime}(\alr(\mathbf{y}_{t-k}) - \boldsymbol{\eta}_{t-k}) \nonumber \\
    &\quad + \mathbf{z}_t^{\prime}\boldsymbol{\gamma},
    \label{phi_t}
\end{align}
where \( \alpha_l \in \mathbb{R} \) are auto-regressive coefficients for the log-precision, \( \tau_k \in \mathbb{R} \) are coefficients associated with past innovations, and \( \mathbf{z}_t \in \mathbb{R}^{r_{\gamma}} \) are known covariates with coefficients \( \boldsymbol{\gamma} \in \mathbb{R}^{r_{\gamma}} \).

For compactness, we write B-DARCH$(P,Q)$ to denote a B-DARMA mean model of order $(P,Q)$ coupled with a DARCH precision process for $\phi_t$ of order $(L,K)$. In general, $(L,K)$ need not equal $(P,Q)$; when they differ, we state both explicitly as B-DARMA$(P,Q)$--DARCH$(L,K)$ (e.g., B-DARMA$(1,0)$--DARCH$(1,1)$). The DARCH component governs time-variation in the precision $\phi_t$, accommodating volatility clustering and covariate effects; see Section~\ref{simulation} for cases with $(L,K)\neq(P,Q)$.

\emph{Alternative log-ratio transformations}, such as the centered log-ratio (CLR) or 
the isometric log-ratio (ILR), can also be used. A detailed discussion of these transformations is provided in the Supplementary Material (Section~\ref{sec:alt-logratios}). 

\subsection{Joint Predictive Distribution}

Define the consecutive observations $\bfy_{a:b} = (\bfy_a, \dots, \bfy_b)^\prime$ for positive integers $a < b$. To be well defined, linear predictor \eqref{eta_f} requires having $m$ previous observations $\bfy_{(t-m):(t-1)}$, 
and corresponding linear predictors $\bfeta_{t-m}, \ldots, \bfeta_{t-1}$. In computing posteriors, we condition on the first $m$ observations $\bfy_{1:m}$ which then do not contribute to the likelihood. For the corresponding first $m$ linear predictors, on the right hand side of \eqref{eta_f}, we set $\bfeta_{1}, \ldots, \bfeta_{m}$ equal to $\alr(\bfy_{1}), \ldots, \alr(\bfy_{m})$ which effectively omits the VMA terms $\pmb{B}_l( \alr(\bfy_{t-l}))$ from \eqref{eta_f} and the MA terms $\tau_l$ $( \alr(\mathbf{y}_{t-l}) - \boldsymbol{\eta}_{t-l})^{\prime}(\alr(\mathbf{y}_{t-l}) - \boldsymbol{\eta}_{t-k})$ from \eqref{phi_t} when $t-l \le m$. In contrast, in \eqref{eta_f} the VAR terms and $\bfX_t\bfbeta$ and in \eqref{phi_t} the AR terms and $\mathbf{z}_t^{\prime}\boldsymbol{\gamma}$ are well defined for $t = 1, \ldots, m$.

Define the $C$-vector $\bftheta$ of all unknown parameters $\bftheta = (\bfA_{prs}, \bfB_{qrs}, \bfbeta^\prime, \bfgamma^\prime,\tau_j,\alpha_i)^\prime$, where $r,s = 1, \ldots, J - 1$ index all elements of matrices $\bfA_p$ and $\bfB_q$, $p = 1, \ldots, P$, $q = 1, \ldots, Q$,  $i=1,\dots, L$, $j=1,\dots,K$  and $C = (P+Q)*(J-1)^2 +L+K + r_\beta + r_\gamma$. Our prior beliefs about $\bftheta$ ($p(\bftheta)$) are updated by Bayes' theorem to give the posterior
\begin{align*}
    p(\bftheta |\bfy_{1:T}) = 
    \frac{p(\bftheta)p(\bfy_{(m+1):T}|\bftheta, \bfy_{1:m})}{p(\bfy_{(m+1):T} | \bfy_{1:m})},
\end{align*}
where $p(\bfy_{(m+1):T}|\bftheta, \bfy_{1:m}) = \prod_{t=m+1}^T p(\bfy_t | \bftheta, \bfy_{(t-m):(t-1)})$, $p(\bfy_t | \bftheta, \bfy_{(t-m):(t-1)})$ is the density of the Dirichlet in \eqref{dirichlet}, and the normalizing constant $p(\bfy_{(m+1):T}| \bfy_{1:m}) = \int p(\bftheta)p(\bfy_{(m+1):T}|\bftheta, \bfy_{1:m}) d\bftheta$. 
Our objective is to forecast the upcoming $S$ observations, denoted as $\mathbf{y}_{(T+1):(T+S)}$. The joint predictive distribution for these future observations is

\begin{align*}
p(\mathbf{y}_{(T+1):(T+S)} \mid \mathbf{y}_{1:T}) = \int_{\boldsymbol{\theta}} p(\mathbf{y}_{(T+1):(T+S)} \mid \boldsymbol{\theta}) \, p(\boldsymbol{\theta} \mid \mathbf{y}_{1:T}) \, d\boldsymbol{\theta}.
\end{align*}

\subsection{Differences between B-DARMA, B-tVARMA, and B-DARCH}

The three models differ in how they respect the simplex and how they treat time‑varying volatility. B‑tVARMA moves the composition to the ALR space and fits a Gaussian VARMA with constant observation covariance unless explicitly extended. This keeps estimation simple but can misalign with the simplex after back‑transformation if the Gaussian–ALR assumptions are strained. B‑DARMA keeps the Dirichlet likelihood on the simplex; its precision $\phi_t$ is a deterministic regression in covariates, so dispersion may change with seasonality or trend but does not react to forecast errors. B‑DARCH also retains the Dirichlet likelihood but replaces the fixed precision with a GARCH‑type, observation‑driven recursion for 
log $\phi_t$. Volatility is updated dynamically from past residuals and past precision, allowing clustering and short‑run bursts in dispersion while preserving the compositional support. This is time‑varying volatility in an observation‑driven sense, not a parameter‑driven stochastic‑volatility state‑space model. In practice, B‑tVARMA is parsimonious but may under‑represent compositional and volatility features; B‑DARMA captures the simplex with covariate‑driven variance; and B‑DARCH adds an adaptive volatility mechanism that improves fit and forecasting when conditional heteroskedasticity matters.

\paragraph{Assumption sensitivity and alternatives.}
Both B\text{-}tVARMA and the Dirichlet-based B\text{-}DARMA/B\text{-}DARCH are parametric choices that can distort if their distributional assumptions are violated. 
On the ALR scale, B\text{-}tVARMA assumes Gaussian residuals; heavy tails or skewness can lead to bias on back-transformation to the simplex. 
For B\text{-}DARMA and B\text{-}DARCH, the Dirichlet likelihood imposes a specific mean–variance relationship and a restricted covariance structure 
(\(\operatorname{Cov}(y_i,y_j)=-\mu_i\mu_j/(\phi_t+1)\) for \(i\neq j\)), which rules out positive contemporaneous correlations among components. 
When the empirical dispersion or dependence deviates from these constraints, Dirichlet-based models can also misrepresent uncertainty or cross-component co-movement.

Other families for proportions are available, including logistic-normal models on a log-ratio scale, generalized Dirichlet specifications, 
mixtures (e.g., Dirichlet mixtures), and zero-augmented variants for sparse compositions. 
We focus on the Dirichlet class because it guarantees simplex-valid predictions and enables a tractable, observation-driven volatility mechanism via the DARCH recursion. 
The transformed-Gaussian comparator (B\text{-}tVARMA) provides a complementary baseline with a richer covariance structure in ALR space. 
In applications where positive cross-correlations or more flexible dispersion are essential, generalized Dirichlet or logistic-normal formulations are natural extensions that can be substituted into the same modeling workflow.

\section{Simulation Study}
\label{simulation}

\subsection{Overview}

We evaluate forecasting performance under two generative mechanisms for compositional time series—Dirichlet ARMA (DARMA) and transformed‑data Gaussian VARMA (tVARMA)—each under two perturbations (misreported observations; temporary regime shifts), yielding four simulation studies in total. In addition to B\text{-}DARMA(1,0), B\text{-}tVARMA(1,0), and the volatility‑aware B\text{-}DARCH(1,0), we also consider a time‑varying‑parameter comparator in ALR space (B\text{-}TVP\text{-}tVARMA) with
\[
\boldsymbol{\eta}_t
=
\sum_{p=1}^{P}\bfA_{t,p}\!\big(\bY_{t-p}-\bX_{t-p}\bfbeta\big)
+
\sum_{q=1}^{Q}\bfB_{t,q}\!\big(\bY_{t-q}-\boldsymbol{\eta}_{t-q}\big)
+
\bX_t\bfbeta,
\qquad
\bY_t\mid\boldsymbol{\eta}_t \sim \Normal(\boldsymbol{\eta}_t,\bfSigma),
\]
and constant observation covariance \(\bfSigma=\diag(\boldsymbol{\sigma})\,\bfOmega\,\diag(\boldsymbol{\sigma})\).
The time‑varying coefficient matrices follow mean‑reverting AR(1) evolutions around long‑run targets,
\[
\bfA_{t,p}
=
\bar{\bfA}_p+\rho^{(A)}_p\!\big(\bfA_{t-1,p}-\bar{\bfA}_p\big)+\tau^{(A)}_{p}\,\bfZ^{(A)}_{t,p},
\qquad
\bfB_{t,q}
=
\bar{\bfB}_q+\rho^{(B)}_q\!\big(\bfB_{t-1,q}-\bar{\bfB}_q\big)+\tau^{(B)}_{q}\,\bfZ^{(B)}_{t,q},
\]
with \(\vecop(\bfZ^{(A)}_{t,p})\) and \(\vecop(\bfZ^{(B)}_{t,q})\stackrel{\text{iid}}{\sim}\Normal(\mathbf{0},\mathbf{I})\).
For the simulation studies we use the B\text{-}TVP\text{-}VAR(1) special case (\(P{=}1\), \(Q{=}0\)) with
\[
\boldsymbol{\eta}_t
=
\bfA_{t,1}\!\big(\bY_{t-1}-\bX_{t-1}\bfbeta\big)+\bX_t\bfbeta,
\qquad
\bfA_{t,1}
=
\bar{\bfA}_1+\rho^{(A)}_1\!\big(\bfA_{t-1,1}-\bar{\bfA}_1\big)+\tau^{(A)}\,\bfZ^{(A)}_{t},
\]
where \(\vecop(\bfZ^{(A)}_{t})\stackrel{\text{iid}}{\sim}\Normal(\mathbf{0},\mathbf{I})\) and \(\tau^{(A)}\sim\HalfNormal(0,\,0.1^2)\).

Across all studies we report out‑of‑sample forecasting accuracy and in‑sample residual autocorrelation.

\subsection{Simulation Setup}
Each dataset contains five compositional components observed over $T=100$ time points; the first 60 observations form the training set and the remaining 40 the test set. For each simulation, parameters are redrawn (intercepts, autoregressive matrices, and—where applicable—precision or covariance) to ensure heterogeneity across runs. Initialization uses a perturbed balanced composition,
\[
y_{1j} = 0.2 + \epsilon_j,\quad \epsilon_j \sim \Normal(0,\,0.01^2),\ j=1,\dots,5,\qquad
y_1 \leftarrow \frac{y_1}{\sum_{j=1}^5 y_{1j}}.
\]
Across all studies, the latent ALR-scale mean follows
\[
\boldsymbol{\eta}_{t} \;=\; \bfbeta \;+\; \bfA\,\big(\mathrm{alr}(\bY_{t-1}) - \bfbeta\big),
\]
with a $4\times4$ matrix $\bfA$ drawn elementwise as $A_{ij}\sim\mathcal{U}(-0.75,\,0.75)$ for each simulation. The simplex intercept is obtained by drawing $\beta_j^\ast \sim \Normal(0.2,\,0.03^2)$ for $j=1,\dots,5$, normalizing to $\sum_j \beta_j^\ast=1$, and applying ALR: $\bfbeta=\mathrm{alr}(\boldsymbol{\beta}^\ast)$.

\paragraph{Comparator defined: B-TVP–VAR(1).}
To allow time‑variation in linear dynamics under constant observation variance, the B‑TVP–VAR(1) comparator specifies
\[
\boldsymbol{\eta}_t \;=\; \bfA_{t,1}\!\big(\bY_{t-1}-\bX_{t-1}\bfbeta\big) \;+\; \bX_t\bfbeta,
\qquad
\bY_t\mid\boldsymbol{\eta}_t \sim \Normal(\boldsymbol{\eta}_t,\,\bfSigma),\quad
\bfSigma=\diag(\boldsymbol{\sigma})\,\bfOmega\,\diag(\boldsymbol{\sigma}).
\]
The coefficient matrix follows a mean‑reverting AR(1) evolution around a long‑run target:
\[
\bfA_{t,1} \;=\; \bar{\bfA}_1 \;+\; \rho^{(A)}_1\!\big(\bfA_{t-1,1}-\bar{\bfA}_1\big) \;+\; \tau^{(A)}\,\bfZ^{(A)}_{t},
\]
with \(\vecop(\bfZ^{(A)}_{t})\stackrel{\text{iid}}{\sim}\Normal(\mathbf{0},\mathbf{I})\) and \(\tau^{(A)}\sim\HalfNormal(0,\,0.1^2)\).

\subsection{Data-Generating Processes (DGPs)}

\subsubsection*{Studies 1 \& 3: DARMA}
Compositions are drawn from a Dirichlet ARMA model with constant precision:
\[
\bY_t \mid \boldsymbol{\mu}_t, \phi_0 \sim \mathrm{Dirichlet}(\phi_0 ,\boldsymbol{\mu}_t),
\qquad
\boldsymbol{\mu}_t \;=\; \mathrm{alr}^{-1}(\boldsymbol{\eta}_t),
\]
with $\phi_0 \sim \mathcal{U}(6,\,7.5)$ independently per simulation.

\subsubsection*{Studies 2 \& 4: tVARMA}
On the ALR scale,
\[
\boldsymbol{\eta}_t \;\sim\; \Normal\!\big(\bfbeta + \bfA(\mathrm{alr}(\bY_{t-1})-\bfbeta),\ \bfSigma\big), 
\qquad \bY_t \;=\; \mathrm{alr}^{-1}(\boldsymbol{\eta}_t).
\]
For each simulation we set $\bfSigma = \mathbf{M}\mathbf{M}^\top$ with $M_{ij}\sim \mathcal{U}(-0.3,\,0.3)$ i.i.d., yielding diverse cross-component dependence.

\subsection{Perturbations}
\paragraph{Misreported observations.}
During the training window, times are selected via cumulative sums of $\mathrm{Poisson}(6)$ draws. At such times $\ell$, replace the composition with a random simplex draw:
\[
y_{\ell j} \sim \mathcal{U}(0,1)\ \text{i.i.d.},\qquad 
\bY_{\text{misreported}[\ell]} \;=\; \frac{(y_{\ell 1},\dots,y_{\ell 5})}{\sum_{j=1}^5 y_{\ell j}}.
\]

\paragraph{Temporary regime shifts.}
Draw a shift time $t_{\text{shift}}\in\{10,\dots,50\}$. For $t\in[t_{\text{shift}},\,t_{\text{shift}}+9]$, redraw parameters—$\bfbeta$, $\bfA$, and the dispersion term for the current DGP (for DARMA, $\phi_0$; for tVARMA, $\bfSigma$). After this 10‑period window, parameters revert to their original values.

\subsection{Fitted Models and Priors}
Each simulated dataset is fit with four Bayesian models: B-DARMA(1,0)-DARCH(1,1) (B-DARCH(1,1)) , B\text{-}DARMA(1,0), B\text{-}tVARMA(1,0), and B-TVP\text{-}VARMA(1,0). All models are implemented in Stan.

\paragraph{Priors.}
Intercepts $\bfbeta \sim \Normal(0,\,0.3^2)$. Elements of $\bfA_p$ in the fixed‑coefficient models receive elementwise $\Normal(0,\,1)$ priors. Gaussian‑error models use
\[
\bfSigma \;=\; \diag(\boldsymbol{\sigma})\,\bfOmega\,\diag(\boldsymbol{\sigma}),
\qquad
L_{\Omega} \sim \LKJChol(3),\quad \sigma_c \sim \HalfNormal(0,\,0.5^2).
\]
For B‑DARCH, the initial log‑precision $\phi_0 \sim \Normal(7,\,1.5^2)$ and the AR/MA terms in $\log \phi_t$ have $\Normal(0.5,\,0.15^2)$ and $\Normal(-0.75,\,0.15^2)$ priors, respectively.

For the B‑TVP–VARMA(1,0) comparator we impose mean reversion on the time‑varying coefficient matrix around a long‑run target. Specifically,
\[
\bfA_{t,1} \;=\; \bar{\bfA}_1 \;+\; \rho^{(A)}_1\!\big(\bfA_{t-1,1}-\bar{\bfA}_1\big) \;+\; \tau^{(A)}\,\bfZ^{(A)}_{t},
\quad
\vecop\!\big(\bfZ^{(A)}_{t}\big)\iid\Normal(\mathbf{0},\mathbf{I}),
\]
with priors
\[
\text{diag}(\bar{\bfA}_1)\sim \Normal(0.5,\,0.3^2),\quad
\bar{\bfA}_1^{(i\ne j)}\sim \Normal(0,\,0.2^2),\quad
\rho^{(A)}_1\sim \BetaDist(9,1),\quad
\tau^{(A)}\sim \HalfNormal(0,\,0.1^2),
\]
and shrinkage of the initial state to the target,
\[
\vecop\!\big(\bfA_{1,1}-\bar{\bfA}_1\big)\sim \Normal(\mathbf{0},\,0.2^2\mathbf{I}).
\]

\paragraph{Computation.}
We run four chains with 2{,}000 iterations per chain (1{,}000 warm‑up), yielding 4{,}000 post‑warm‑up draws per model. Convergence diagnostics and runtimes are reported in the supplement.

\subsection{Metrics}
We assess model performance using \textbf{Forecast Root Mean Squared Error (FRMSE)} and \textbf{Forecast Mean Absolute Error (FMAE)} for each of the five components. These metrics quantify the out-of-sample forecast accuracy
    \[
    \FRMSE_j = \left( \frac{1}{40} \sum_{t = 61}^{100} \left( y_{tj} - \bar{\mu}_{tj} \right)^2 \right)^{\frac{1}{2}}
    \]
    \[
    \FMAE_j = \frac{1}{40} \sum_{t = 61}^{100} \left| y_{tj} - \bar{\mu}_{tj} \right|
    \]
    where \( \bar{\mu}_{tj} \) represents the posterior mean of the predicted value at time $t$ for component \( j \). We then summarize by averaging them across all five components $
    \FRMSE_{mean} = \frac{1}{5}\sum_{j=1}^5 \FRMSE_j $ and $\FMAE_{mean} =\frac{1}{5}\sum_{j=1}^5 \FMAE_j $.

Additionally, we assess the \textbf{Partial Autocorrelation Function (PACF)} (see Supplementary Section~\ref{sec:pacf}) of the sums of squared standardized residuals to evaluate how well each model captures residual dynamics during the in-sample period. For each model, residuals are standardized according to the model's underlying distribution (see Supplementary Section~\ref{sec:ssr}).

\subsection{Results}

\subsubsection{Simulation Studies 1--2: Shocks}
Table~\ref{tab:model_performance_sims12} summarizes out-of-sample performance under
(i) a DARMA DGP with misreported observations (Sim~1) and
(ii) a tVARMA DGP with misreported observations (Sim~2).
In both designs, B\text{-}DARCH attains the lowest average FRMSE and FMAE among all competitors.
Relative to the second-best model, the FRMSE improvement is about $1.3\%$ in Sim~1
and $1.1\%$ in Sim~2; the corresponding FMAE improvements are about $2.7\%$ and $2.4\%$.
The B-TVP-tVARMA comparator tracks B\text{-}tVARMA closely but remains slightly behind B\text{-}DARCH
on both metrics.

Figure~\ref{fig:pacf_sim_shock} displays mean PACF values of the sums of squared standardized residuals across replications.
Under shock contamination, all models exhibit modest negative partial autocorrelation at the first few lags,
consistent with one-step “correction” after a misreport.
B\text{-}DARCH’s bars lie closest to zero for lags 1--6 and decay rapidly thereafter,
indicating the least residual short-run dependence among the four methods.

\subsubsection{Simulation Studies 3--4: Regime Shifts}
Table~\ref{tab:model_performance_sims34} reports results when the DGP undergoes a temporary regime shift:
(iii) DARMA with a transient parameter change (Sim~3) and
(iv) tVARMA with a transient parameter change (Sim~4).
B\text{-}DARCH again has the smallest average FRMSE and FMAE in both settings.
The gains over the second-best model are modest but consistent:
about $1.0\%$ (FRMSE) and $2.3\%$ (FMAE) in Sim~3, and
about $0.2\%$ (FRMSE) and $0.9\%$ (FMAE) in Sim~4.
Under a DARMA regime shift (Sim~3) the B-TVP-tVARMA comparator is the clear runner-up,
whereas under a tVARMA regime shift (Sim~4) B\text{-}tVARMA is the runner-up.

Figure~\ref{fig:pacf_sim_regime} shows the corresponding mean PACF patterns.
With regime shifts, all models display a strong lag-1 spike; for both DGPs the spike is largest for the B-TVP-tVAR class
(B\text{-}DARCH and B-TVP-tVARMA).
Crucially, B\text{-}DARCH decays most rapidly and stays comparatively close to zero beyond lag~3,
while B\text{-}tVARMA and B-TVP-tVARMA exhibit more persistent negative partial autocorrelations
at medium lags (roughly 8--16), indicating longer-lived residual structure once the shift occurs.
Taken together, these diagnostics suggest that B\text{-}DARCH concentrates any remaining dependence immediately after the shift
and then dissipates it faster, whereas fixed-variance comparators leave more medium-horizon autocorrelation.

\paragraph{Summary.}
Across all four designs (DARMA/tVARMA $\times$ shocks/regime shifts), B\text{-}DARCH yields the best point-forecast
accuracy (FRMSE/FMAE) and the least residual persistence beyond very short lags.
Because the DGPs considered here have constant variance, the improvements are necessarily modest;
nevertheless, the volatility-aware specification helps the model absorb shocks and adapt quickly to transient parameter changes.
Meanwhile, the B-TVP-tVARMA comparator is competitive, especially under regime shifts, but does not close the gap to B\text{-}DARCH.

\begin{table}[ht]
\centering
\begin{tabular}{|l|cc|cc|}
\hline
\textbf{Model} & \multicolumn{2}{c|}{\textbf{Sim 1 (DARMA Shock)}} & \multicolumn{2}{c|}{\textbf{Sim 2 (tVARMA Shock)}} \\
\hline
 & \textbf{Average FRMSE} & \textbf{Average FMAE} & \textbf{Average FRMSE} & \textbf{Average FMAE} \\
\hline
B-DARMA  & 20.12 & 3.13 & 18.36 & 2.61 \\
B-DARCH  & 19.67 & 2.98 & 18.08 & 2.53 \\
B-tVARMA & 19.97 & 3.07 & 18.28 & 2.59 \\
B-TVP-tVARMA & 19.92 & 3.06 & 18.33 & 2.61 \\
\hline
\end{tabular}
\caption{Summary of model performance metrics ($\times 100$) on the test set ($T{=}40$) across 50 simulations for Simulation Studies 1--2. 
FRMSE ($\times 100$) equals the across-simulation average of 
$\mathrm{RMSE}=\sqrt{\frac{1}{TJ}\sum_{t=1}^{T}\sum_{j=1}^{J}(y_{tj}-\hat y_{tj})^{2}}$; 
FMAE ($\times 100$) equals the across-simulation average of 
$\mathrm{MAE}=\frac{1}{TJ}\sum_{t=1}^{T}\sum_{j=1}^{J}|y_{tj}-\hat y_{tj}|$.}
\label{tab:model_performance_sims12}
\end{table}

\begin{table}[ht]
\centering
\begin{tabular}{|l|cc|cc|}
\hline
\textbf{Model} & \multicolumn{2}{c|}{\textbf{Sim 3 (DARMA Regime Shift)}} & \multicolumn{2}{c|}{\textbf{Sim 4 (tVARMA Regime Shift)}} \\
\hline
 & \textbf{Average FRMSE} & \textbf{Average FMAE} & \textbf{Average FRMSE} & \textbf{Average FMAE} \\
\hline
B-DARMA  & 18.68 & 2.72 & 18.11 & 2.48 \\
B-DARCH  & 18.30 & 2.59 & 17.46 & 2.29 \\
B-tVARMA & 18.65 & 2.73 & 17.49 & 2.31 \\
B-TVP-VAR & 18.49 & 2.65 & 18.03 & 2.45 \\
\hline
\end{tabular}
\caption{Summary of model performance metrics ($\times 100$) on the test set ($T{=}40$) across 50 simulations for Simulation Studies 3--4. 
FRMSE ($\times 100$) equals the across-simulation average of 
$\mathrm{RMSE}=\sqrt{\frac{1}{TJ}\sum_{t=1}^{T}\sum_{j=1}^{J}(y_{tj}-\hat y_{tj})^{2}}$; 
FMAE ($\times 100$) equals the across-simulation average of 
$\mathrm{MAE}=\frac{1}{TJ}\sum_{t=1}^{T}\sum_{j=1}^{J}|y_{tj}-\hat y_{tj}|$.}
\label{tab:model_performance_sims34}
\end{table}

\section{Airbnb Data Analysis}
\label{airbnb}
The dataset comprises daily records from January 1, 2017, to December 31, 2020, detailing the proportion of fees in U.S. Dollars (USD) originating from various billing currencies. The data is categorized into four anonymized diverse geographic regions, referred to as Regions 1-4. Within each region, the fees are broken down by the top five currencies specific to that region, with the remaining currencies grouped into an ``other'' category. The top five currencies for each region are fixed for the entire period. The training period spans from January 1, 2017, to June 30, 2020. The subsequent three months (July 1, 2020 to September 30, 2020) are used for validation and model specification, while the final 3 months (October 1, 2020, to December 31, 2020) constitute the test set.

\subsection{Exploratory Data Analysis}

\subsubsection{Regional Currency Dynamics}

Three features of the data motivate our modeling choices: pronounced seasonality, strong temporal dependence, and recurrent volatility shifts.

\begin{figure}
    \includegraphics[scale=.75]{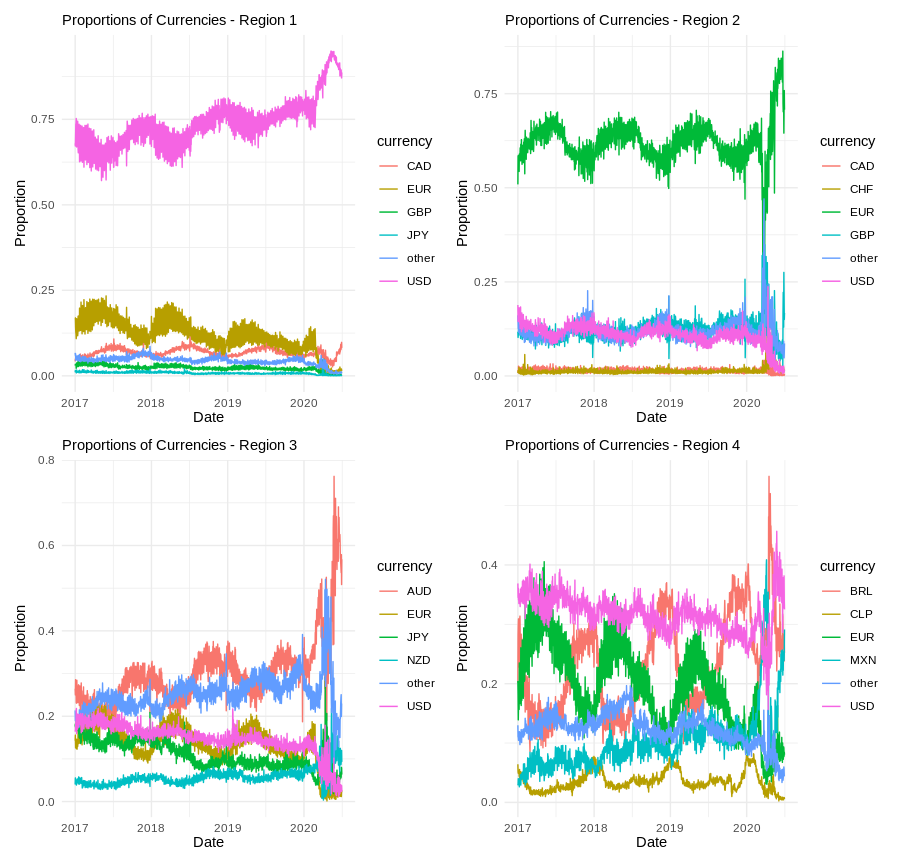}
    \caption{Airbnb data analysis - proportion of fees by billing currency for four regions from Jan 1, 2017 to June 30, 2020. AUD is the Australian dollar, BRL is the Brazillian Real, CAD is the Canadian Dollar,CHF is the Swiss Franc, CLP is the Chilean Peso,EUR is the European Euro, GBP is the Great British Pound, MXN is the Mexican Peso, NZD is the New Zealand Dollar, and USD is the US Dollar. }
    \label{training_all}
\end{figure}

\paragraph{Seasonality and trends.}
Figure~\ref{training_all} displays currency proportions across the four regions. All regions exhibit clear yearly seasonal cycles, with Region~4 showing the strongest patterns. Pre-pandemic trends vary by region: Regions~1 and~2 show steady or declining currency shares relative to USD, while Region~3 exhibits upward drift in AUD. The COVID-19 period (early 2020) induced sharp, region-specific disruptions. USD surged in Region~1, EUR and CHF rose in Region~2, and BRL experienced large swings in Region~4. These trend breaks motivate our inclusion of flexible seasonal components and suggest the need for models that are robust to structural shifts.

\paragraph{Temporal dependence.}
Lag-1 autocorrelations of ALR-transformed series are uniformly high across regions (Supplementary Figure~\ref{alr_correlation}), confirming that currency compositions are persistent and justifying autoregressive structure in our specifications.

\begin{figure}
    \includegraphics[scale=.75]{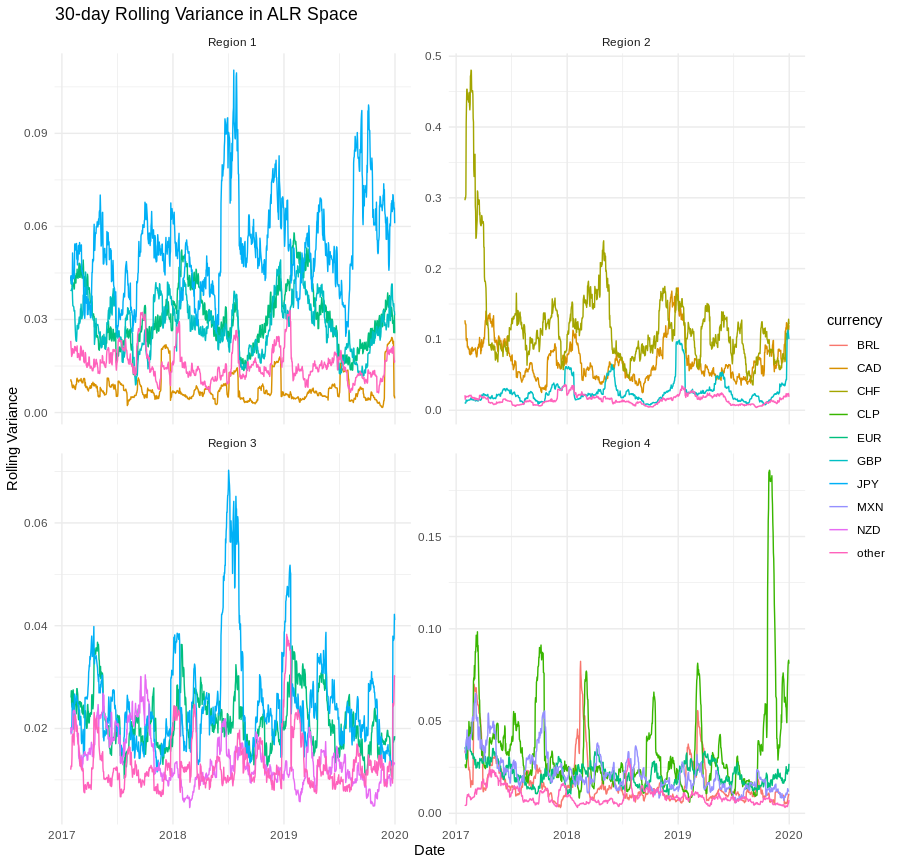}
    \caption{Airbnb data analysis - proportion of fees by currency- 30-day rolling ALR variance for four regions from Jan 1, 2017 to Dec 31, 2019. USD is the reference component.}
    \label{covid_alr_variances}
\end{figure}

\paragraph{Time-varying volatility.}
Figure~\ref{covid_alr_variances} reveals that volatility in ALR-transformed proportions is not constant but exhibits recurrent episodes of elevation. Rather than a single shock, we observe multiple surges across all four regions, often persisting for weeks before subsiding. Region~2 begins with elevated variance in CAD and CHF, settles to moderate levels, then spikes again in early 2020. Region~4 shows abrupt climbs in BRL and CLP near 2019 with additional spikes emerging after 2020. These patterns indicate that volatility is shaped by evolving market conditions rather than one-off events. This observation is the central motivation for our DARCH specification.

\clearpage

\subsection{Objective}
The primary goal is to compare \emph{four} model classes for forecasting currency–share compositions:
(i) Bayesian Dirichlet ARMA (B\text{-}DARMA),
(ii) Bayesian Dirichlet ARMA with dynamic precision (B\text{-}DARMA\text{-}DARCH; “B\text{-}DARCH”),
(iii) Bayesian transformed Gaussian VARMA (B\text{-}tVARMA), and
(iv) a time\text{-}varying\text{-}parameter comparator in ALR space (B\text{-}TVP\text{-}tVARMA).
All models are evaluated under a common training/validation/test split with weekly and yearly seasonality and a linear trend included in the design.

\subsection{Comparison Models}

\paragraph{B\text{-}DARCH (Dirichlet with DARCH precision).}
B\text{-}DARCH and B\text{-}DARMA share the same ALR\text{-}scale mean model for \(\bfeta_t\), but B\text{-}DARCH endows the Dirichlet precision \(\phi_t\) with its own AR/MA recursion on \(\log\phi_t\), allowing time\text{-}varying dispersion (time‑varying volatility (conditional heteroskedasticity)) while keeping predictions on the simplex.

\paragraph{B\text{-}DARMA (Dirichlet with fixed precision regression).}
Identical mean specification to B\text{-}DARCH, but the precision is deterministic given covariates \(\bfz_t\) (no time‑varying volatility (conditional heteroskedasticity)). Temporal dependence is handled in the mean only.

\paragraph{B\text{-}tVARMA (transformed\text{-}Gaussian benchmark).}
Transform the composition via \(\bY_t=\alr(\bfy_t)\in\mathbb{R}^{J-1}\) and fit
\[
\bY_t \sim \Normal(\boldsymbol{\eta}_t,\ \bfSigma),\qquad
\boldsymbol{\eta}_t
= \sum_{p=1}^{P}\bfA_p\!\big(\bY_{t-p}-\bX_{t-p}\bfbeta\big)
+ \sum_{q=1}^{Q}\bfB_q\!\big(\bY_{t-q}-\boldsymbol{\eta}_{t-q}\big)
+ \bX_t\bfbeta,
\]
with \(\bfSigma\) unknown. This model flexibly captures mean dynamics in ALR space but assumes time\text{-}invariant observation variance.

\paragraph{B\text{-}TVP\text{-}tVARMA (time-varying-parameter comparator in ALR space).}
On the ALR scale we take
\[
\bY_t \mid \boldsymbol{\eta}_t \sim \Normal(\boldsymbol{\eta}_t,\ \bfSigma),
\qquad
\bfSigma=\diag(\boldsymbol{\sigma})\,\bfOmega\,\diag(\boldsymbol{\sigma}).
\]
The conditional mean includes both VAR and VMA components,
\[
\boldsymbol{\eta}_t
=
\sum_{p=1}^{P}\bfA_{t,p}\big(\bY_{t-p}-\bX_{t-p}\bfbeta\big)
+
\sum_{q=1}^{Q}\bfB_{t,q}\big(\bY_{t-q}-\boldsymbol{\eta}_{t-q}\big)
+
\bX_t\bfbeta .
\]
Coefficient matrices evolve by mean reversion toward long-run targets,
\[
\bfA_{t,p}=\bar{\bfA}_p+\rho^{(A)}_{p}\big(\bfA_{t-1,p}-\bar{\bfA}_p\big)+\tau^{(A)}_{p}\,\bfZ^{(A)}_{t,p},
\quad
\bfB_{t,q}=\bar{\bfB}_q+\rho^{(B)}_{q}\big(\bfB_{t-1,q}-\bar{\bfB}_q\big)+\tau^{(B)}_{q}\,\bfZ^{(B)}_{t,q},
\]
with \(\vecop(\bfZ^{(A)}_{t,p})\) and \(\vecop(\bfZ^{(B)}_{t,q})\) i.i.d. \(\Normal(\mathbf{0},\mathbf{I})\).

\paragraph{Computation.}
All four models were fit in Stan with 4 chains (2{,}000 iterations; 1{,}000 warm‑up). 
Diagnostics were satisfactory across regions and models (no divergences; only rare max tree‑depth hits; 
worst $\hat{R}=1.01$; strong bulk/tail ESS); see Table~\ref{tab:nuts}.

\subsection{Seasonal and Trend Terms}
Figure \ref{training_all} and Supplementary figure \ref{daily_seasonal} show varying levels of yearly and weekly seasonality in the currency data in different regions. There is a recurring yearly pattern driven by popular travel periods and a clear weekly seasonality reflecting variations between days of the week. Consequently, we include weekly and yearly seasonal variables and a trend variable in the predictors $\mathbf{X}_t$ and the same weekly and yearly seasonal variables in $\bfz_t$. Both $\phi_t$ and the elements of $\bfeta_t$ are modeled with a linear trend and Fourier terms for seasonal variables, using pairs of $\left(\sin{\frac{2 k \pi t}{w_{\text{season}}}}, \cos{\frac{2 k \pi t}{w_{\text{season}}}}\right)$ for $k = 1, \dots, K_{\text{season}} \leq \frac{w_{\text{season}}}{2}$, with $w_{\text{season}} = 7$ for weekly seasonality and $w_{\text{season}} = 365.25$ for yearly seasonality. The orthogonality of the Fourier terms aids in model convergence. The B-DARCH, B-DARMA, B-tVARMA, and B-TVP-tVARMA all share the same $\mathbf{X}_t$ while the B-DARCH and B-DARMA share the same $\bfz_t$.

\subsection{Model Specification and Validation}
We train on January~1, 2017 to June~30, 2020; July~1 to September~30, 2020 serves as a validation window for choosing lags and seasonal complexity; October~1 to December~31, 2020 is the held\text{-}out test set. For the initial pass we set \(P=1\), \(Q=0\), and \(K_{\text{week}}=3\) for each region, then select the optimal number of yearly Fourier terms \(K_{\text{year}}\) by validation performance. With \(K_{\text{year}}\) fixed, we vary \(P\) and \(Q\) on the validation set to choose AR/MA orders. Using the chosen specifications, all four models (B\text{-}DARMA, B\text{-}DARCH, B\text{-}tVARMA, and B\text{-}TVP\text{-}tVARMA) are refit through September~30, 2020 and evaluated on the independent test set.

\subsection{Priors}

\paragraph{Shared structure.}
Across models we encode the belief that own–lag effects dominate by assigning diagonal AR/MA entries $(a_{rrp},b_{rrq})\sim\Normal(0.4,\,0.5^2)$ and off–diagonals $\sim\Normal(0,\,0.5^2)$. Seasonal Fourier terms, linear trends, and intercepts use $\Normal(0,1)$, $\Normal(0,0.1^2)$, and $\Normal(0,2^2)$, respectively. For Dirichlet models, the precision–side design $\bfz_t$ uses the same seasonal basis as the mean design $\bX_t$.

\paragraph{B\text{-}DARMA and B\text{-}DARCH.}
In both Dirichlet models the precision covariates $\bfz_t$ adopt the same priors as on the mean side. For B\text{-}DARCH, the AR and MA coefficients in $\log\phi_t$ each have $\Normal(0,1)$ priors, and the initial log‑precision is given a weakly informative normal prior.

\paragraph{B\text{-}tVARMA.}
On the ALR scale we take
\[
\boldsymbol{\Sigma}
= 
\operatorname{diag}(\boldsymbol{\sigma})\,\boldsymbol{\Omega}\,\operatorname{diag}(\boldsymbol{\sigma}),
\qquad
\sigma_c \sim \HalfNormal(0.5), \;
L_{\boldsymbol{\Omega}} \sim \LKJChol(3),
\]
yielding an unrestricted yet regularized covariance for the transformed-Gaussian likelihood.

\paragraph{B--TVP--tVARMA.}
The TVP comparator uses the same ALR‑scale likelihood $\bY_t\mid\bfeta_t\sim\Normal(\bfeta_t,\bfSigma)$ but with $\bfSigma=\diag(\boldsymbol{\sigma})\,\bfOmega\,\diag(\boldsymbol{\sigma})$, where each component $\sigma_c\sim\HalfNormal(0,0.5^2)$ and $L_\Omega\sim\LKJChol(3)$. Time‑varying coefficient matrices follow mean‑reverting AR(1) evolutions around long‑run targets: $\bfA_{t,p}=\bar{\bfA}_p+\rho^{(A)}_{p}(\bfA_{t-1,p}-\bar{\bfA}_p)+\tau^{(A)}_{p}\bfZ^{(A)}_{t,p}$ and analogously for $\bfB_{t,q}$, with $\vecop(\bfZ)\iid\Normal(\mathbf{0},\mathbf{I})$, $\rho^{(A)}_{p},\rho^{(B)}_{q}\sim\BetaDist(9,1)$, and $\tau^{(A)}_{p},\tau^{(B)}_{q}\sim\HalfNormal(0,.1^2)$. Long‑run targets are weakly centered by taking diagonal elements $\Normal(0.5,0.3)$ and off‑diagonals $\Normal(0,0.2)$, and initial states shrink to these targets with $\Normal(0,0.2)$ per entry. Exogenous effects are shared across models via $\bfbeta\sim\Normal(\mathbf{0},\beta_{\!sd}^2\mathbf{I})$.

\subsection{Results}

\subsubsection{Model Specification}

We selected region-specific lag orders and Fourier harmonics via out-of-sample validation, with full results reported in Supplementary Table~\ref{tab:combined_validation}. Optimal specifications varied across regions: Regions~2 and~3 favored richer seasonal structure ($K_{\text{year}}=10$) with ARMA(1,1) dynamics, Region~1 performed best with fewer harmonics ($K_{\text{year}}=8$) and a pure autoregressive specification, while Region~4 required the most complex short-run dynamics at ARMA(3,2) despite fewer harmonics. These differences likely reflect heterogeneity in seasonal persistence and short-run adjustment across geographic markets.

\paragraph{Aggregate performance.}

\begin{figure}
    \includegraphics[scale=.85]{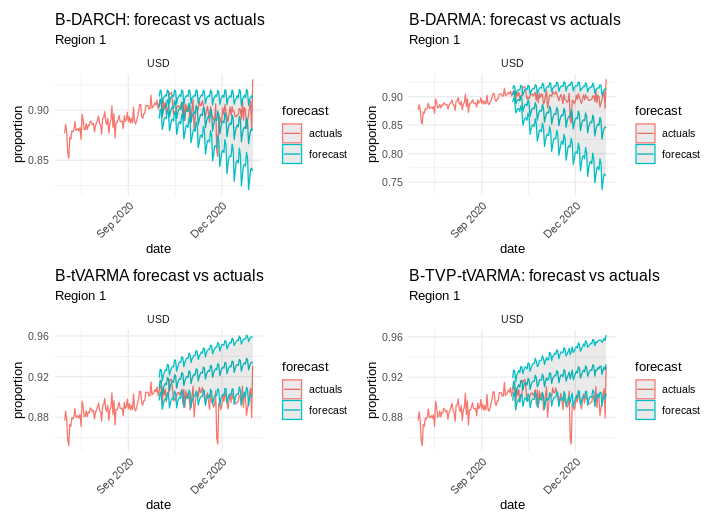}
    \caption{Airbnb data analysis — Region 1: 92-day forecasts (blue) with 95\% \emph{credible} intervals (shaded) for USD currency from Oct.\ 1 to Dec.\ 31, 2020, compared to actual values (red) from the preceding six months. Models shown: B-DARCH, B-DARMA, B-tVARMA, and B-TVP-tVARMA.}
    \label{forecast_region_1}
\end{figure}

\begin{figure}
    \includegraphics[scale=.85]{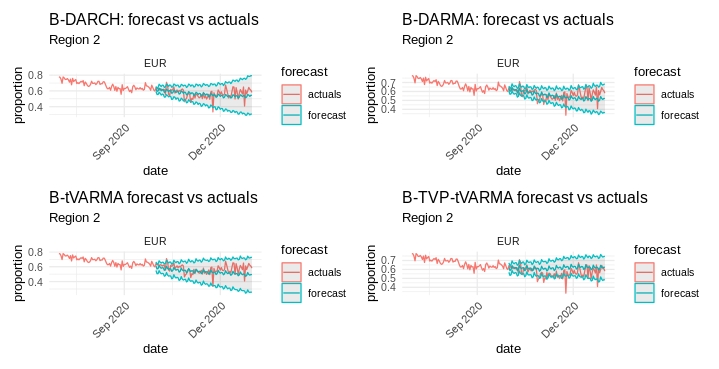}
    \caption{Airbnb data analysis — Region 2: 92-day forecasts (blue) with 95\% \emph{credible} intervals (shaded) for EUR currency from Oct.\ 1 to Dec.\ 31, 2020, compared to actual values (red) from the preceding six months. Models shown: B-DARCH, B-DARMA, B-tVARMA, and B-TVP-tVARMA.}
    \label{forecast_region_2}
\end{figure}

\begin{figure}
    \includegraphics[scale=.85]{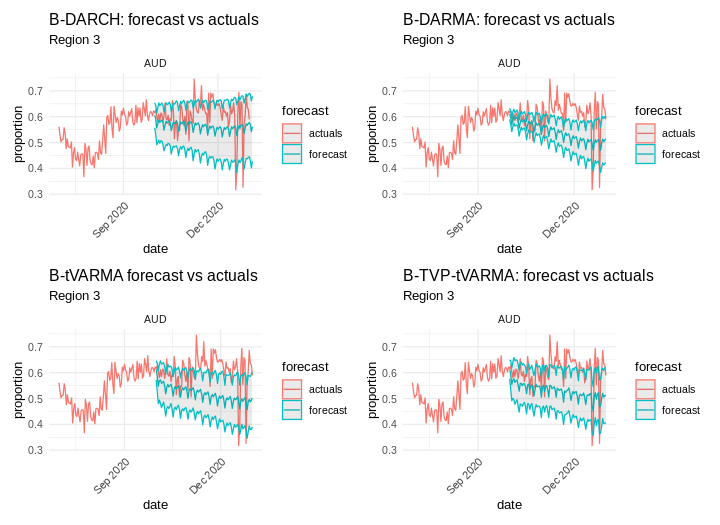}
    \caption{Airbnb data analysis — Region 3: 92-day forecasts (blue) with 95\% \emph{credible} intervals (shaded) for AUD currency from Oct.\ 1 to Dec.\ 31, 2020, compared to actual values (red) from the preceding six months. Models shown: B-DARCH, B-DARMA, B-tVARMA, and B-TVP-tVARMA.}
    \label{forecast_region_3}
\end{figure}

\begin{figure}
    \includegraphics[scale=.85]{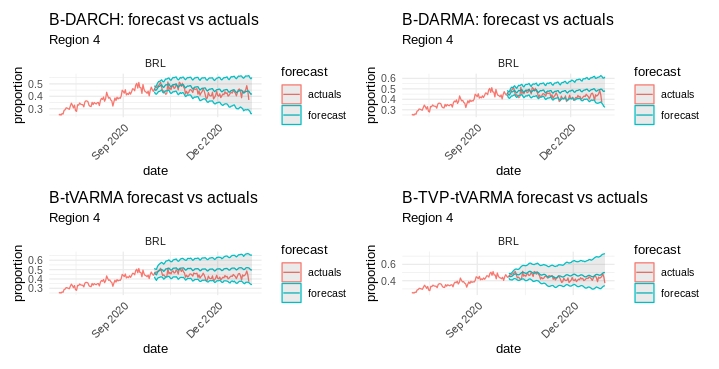}
    \caption{Airbnb data analysis — Region 4: 92-day forecasts (blue) with 95\% \emph{credible} intervals (shaded) for BRL currency from Oct.\ 1 to Dec.\ 31, 2020, compared to actual values (red) from the preceding six months. Models shown: B-DARCH, B-DARMA, B-tVARMA, and B-TVP-tVARMA.}
    \label{forecast_region_4}
\end{figure}

Table~\ref{tab:all_regions_fmae_frss} compares forecast accuracy across four models: B-DARCH, B-DARMA, B-tVARMA, and a time-varying-parameter variant (TVP-tVARMA). Corresponding forecast trajectories for each region's dominant currency appear in Figures~\ref{forecast_region_1}–\ref{forecast_region_4}.

The central finding is that B-DARCH achieves the lowest aggregate FMAE and FRSS in all four regions. The gains are most pronounced in Region~1, where B-DARCH reduces total FMAE by over 60\% relative to B-DARMA and FRSS by more than 80\%. Regions~2–4 show more modest but consistent improvements. These results suggest that explicitly modeling time-varying volatility---the distinguishing feature of the DARCH specification---yields meaningful forecast gains, particularly in markets where volatility shifts are most pronounced.

Among the comparators, TVP-tVARMA generally ranks second, indicating that time-varying parameters provide some benefit even without the DARCH volatility structure. However, this flexibility comes at computational cost: TVP-tVARMA required the longest runtimes, while B-DARMA and B-tVARMA typically completed in approximately one hour (Table~\ref{tab:compute_time}).

\paragraph{Per-currency comparisons.}

While B-DARCH dominates at the aggregate level, currency-level rankings are more nuanced (Table~\ref{tab:all_regions_fmae_frss}). B-DARCH achieves the lowest FMAE for the majority of individual currencies in each region---for instance, five of six currencies in Region~2. However, TVP-tVARMA occasionally outperforms for specific series, including GBP in Region~2, JPY and NZD in Region~3, and MXN and USD in Region~4. In Region~1, B-DARCH leads for CAD, USD, and the residual category, while TVP-tVARMA edges ahead for EUR, JPY, and GBP. B-DARMA is occasionally competitive at the currency level (e.g., USD in Region~3, EUR in Region~4) but never leads in aggregate.

These patterns suggest that DARCH's volatility modeling provides broad benefits across currency compositions, while time-varying mean parameters may better capture dynamics for particular series. Readers interested in specific currencies can consult the detailed breakdown in Table~\ref{tab:all_regions_fmae_frss}.

\clearpage

\begin{table}[htbp]
\centering
\caption{Forecast Mean Absolute Error (FMAE) and Forecast Residual Sum of Squares (FRSS) by region, currency, and model (values $\times 10^2$)}
\label{tab:all_regions_fmae_frss}
\begingroup
\scriptsize                                
\setlength{\tabcolsep}{3pt}                
\renewcommand{\arraystretch}{0.75}  
\begin{tabular}{lrrrrrrrrr} 
\hline
Region & Currency & \multicolumn{4}{c}{FMAE} & \multicolumn{4}{c}{FRSS} \\
       &          & \textbf{DARCH} & \textbf{DARMA} & \textbf{tVARMA} & \textbf{TVP-tVARMA}
                  & \textbf{DARCH} & \textbf{DARMA} & \textbf{tVARMA} & \textbf{TVP-tVARMA} \\
\hline

1 & CAD   & 0.55 & 0.87 & 1.21 & 1.05 & 0.49 & 1.07 & 1.75 & 1.34 \\
1 & EUR   & 0.61 & 2.19 & 0.55 & 0.48 & 0.53 & 5.23 & 0.55 & 0.42 \\
1 & GBP   & 0.18 & 0.43 & 0.21 & 0.18 & 0.05 & 0.22 & 0.06 & 0.04 \\
1 & JPY   & 0.08 & 0.18 & 0.06 & 0.05 & 0.01 & 0.04 & 0.01 & 0.00 \\
1 & USD   & 1.08 & 3.13 & 2.52 & 2.21 & 1.91 &11.70 & 7.65 & 5.85 \\
1 & other & 0.33 & 0.85 & 0.66 & 0.57 & 0.17 & 0.85 & 0.60 & 0.46 \\
\hline
1 & \textbf{total}   & 2.84 & 7.65 & 5.20 & 4.55 & 3.15 & 19.11 & 10.62 & 8.12 \\
1 & \textbf{average} & 0.47 & 1.27 & 0.87 & 0.76 & 0.53 &  3.19 &  1.77 & 1.35 \\
\hline

2 & CAD   & 0.35 & 0.38 & 0.37 & 0.46 &  0.22 &  0.24 &  0.27 &  0.36 \\
2 & CHF   & 0.83 & 0.89 & 0.96 & 1.19 &  0.97 &  1.14 &  1.36 &  2.00 \\
2 & EUR   & 4.51 & 4.85 & 5.51 & 5.65 & 29.70 & 34.50 & 42.20 & 49.70 \\
2 & GBP   & 4.06 & 4.13 & 5.32 & 2.83 & 23.00 & 24.00 & 35.30 & 14.90 \\
2 & USD   & 1.28 & 1.68 & 1.68 & 1.76 &  3.04 &  4.59 &  4.60 &  4.93 \\
2 & other & 2.52 & 2.78 & 3.29 & 2.79 & 17.00 & 17.40 & 20.70 & 18.60 \\
\hline
2 & \textbf{total}   & 13.55 & 14.71 & 17.13 & 14.68 &  73.93 &  81.87 & 104.43 &  90.49 \\
2 & \textbf{average} &  2.26 &  2.45 &  2.86 &  2.45 &  12.32 &  13.65 &  17.41 &  15.08 \\
\hline

3 & AUD   & 6.31 & 8.26 & 9.98 & 8.73 &  57.70 &  88.70 & 116.00 &  89.00 \\
3 & EUR   & 0.64 & 0.93 & 2.40 & 2.10 &   0.53 &   1.10 &   6.03 &   4.62 \\
3 & JPY   & 2.20 & 2.18 & 2.33 & 2.04 &  23.70 &  24.10 &  23.00 &  17.60 \\
3 & NZD   & 3.57 & 2.92 & 1.37 & 1.20 &  14.60 &   9.98 &   2.75 &   2.11 \\
3 & USD   & 0.82 & 0.79 & 2.48 & 2.17 &   1.00 &   0.98 &   6.61 &   5.06 \\
3 & other & 2.69 & 5.09 & 5.14 & 4.49 &  13.30 &  34.80 &  32.70 &  25.00 \\
\hline
3 & \textbf{total}   & 16.23 & 20.17 & 23.70 & 20.73 & 110.83 & 159.66 & 187.09 & 143.39 \\
3 & \textbf{average} &  2.71 &  3.36 &  3.95 &  3.46 &  18.47 &  26.61 &  31.18 &  23.90 \\
\hline

4 & BRL   & 2.84 & 4.36 & 6.38 & 3.42 & 10.80 & 23.60 & 46.20 & 15.80 \\
4 & CLP   & 2.32 & 2.35 & 3.09 & 2.46 &  7.76 &  8.45 & 13.00 &  9.78 \\
4 & EUR   & 1.29 & 0.58 & 0.64 & 1.88 &  2.08 &  0.51 &  0.60 &  4.15 \\
4 & MXN   & 1.27 & 2.85 & 1.54 & 1.15 &  2.36 &  8.71 &  2.98 &  1.98 \\
4 & USD   & 2.70 & 4.14 & 2.94 & 2.35 &  8.68 & 19.60 &  9.93 &  6.93 \\
4 & other & 0.79 & 0.97 & 1.94 & 0.96 &  0.87 &  1.46 &  4.93 &  1.21 \\
\hline
4 & \textbf{total}   & 11.21 & 15.25 & 16.53 & 12.22 & 32.55 & 62.33 & 77.64 & 39.85 \\
4 & \textbf{average} &  1.87 &  2.54 &  2.76 &  2.04 &  5.43 & 10.39 & 12.94 &  6.64 \\
\hline
\end{tabular}
\caption{Airbnb data analysis: Forecast Mean Absolute Error (FMAE) and Forecast Residual Sum of Squares (FRSS) for Regions 1 to 4 and top currencies. FMAE and FRSS values are multiplied by $10^{2}$.}
\label{tab:combined_region_metrics}
\endgroup
\end{table}

\clearpage

\subsubsection{Residual Analysis}
We compute the partial autocorrelation function (PACF) of the sum of standardized squared residuals (SSR) on the test set for each region and model (Supplementary Figures~\ref{pacf_region_1}–\ref{pacf_region_4}). All four models show a prominent spike at lag~1, but its magnitude is systematically smaller under B-DARCH. Using a practical significance threshold of $0.2$, B-DARCH exhibits only isolated exceedances beyond lag~1, indicating that short‑run volatility dynamics have largely been captured. By contrast, B-DARMA, B-tVARMA, and B-TVP-tVARMA display multiple exceedances between lags~2 and~10 in several regions, suggesting remaining residual dependence at intermediate lags. This pattern is consistent with the design of the DARCH component, which allows volatility to adapt dynamically to recent shocks and past variability, thereby reducing serial structure in SSR.

Supplementary Table~\ref{tab:coverage_all_regions} summarizes empirical coverage of 95\% credible intervals. Mean coverage for B-DARCH remains close to nominal across regions (0.92, 0.91, 0.87, 0.91 in Regions 1–4). B-TVP-tVARMA’s coverage is mixed—strong in Region~4 (0.91) but under-covering in Region~1 (0.65) and Region~2 (0.75)—while B-DARMA varies widely and B-tVARMA often undercovers. Together with the forecast metrics, these results highlight that explicitly modeling time‑varying volatility improves both point accuracy and uncertainty calibration, especially in regions with pronounced volatility shifts.

\subsubsection{Parameter Estimates.}
A complete discussion of the posterior densities for the DARCH parameters
(auto-regressive \(\alpha\) and moving average \(\tau\) coefficients) in each 
region appears in the Supplementary Material (Section~\ref{sec:parameter}). 
Briefly, Regions 2 and 3 exhibit higher \(\alpha\) values indicating 
more persistent volatility, while Region 1 shows moderate persistence 
(\(0.25 < \alpha < 0.75\)), and Region 4 displays a more complex 
multi-lag structure. These region-specific patterns underscore 
the flexibility of our B-DARCH approach to accommodate 
various volatility regimes.

\section{Discussion} 
\label{discussion}

We introduced a Bayesian Dirichlet ARMA model with a Dirichlet ARCH-type volatility component (B\text{-}DARCH) for forecasting compositional time series. The key design choice is to keep the likelihood on the simplex and place dynamics not only on the ALR-scale mean but also on the Dirichlet precision, so that dispersion adapts to recent shocks. Alongside standard Dirichlet and transformed-Gaussian benchmarks, we evaluated a time-varying-parameter comparator in ALR space (B\text{-}TVP\text{-}tVARMA). This fourth comparator allows the \emph{mean} dynamics to drift while maintaining constant observation covariance on the ALR scale. 

The simulation evidence is deliberately simple: two data-generating mechanisms (DARMA and tVARMA), each perturbed by misreported observations and temporary regime shifts. In all four designs the observation variance is constant; gains from modeling dynamic precision should therefore be modest if the comparator class is well specified. That is exactly what we find. B\text{-}DARCH achieves the best out-of-sample accuracy on average and exhibits the weakest medium-lag structure in the PACF of squared standardized residuals. The B\text{-}TVP\text{-}tVARMA comparator is competitive, often the runner-up when the \emph{center} drifts under a regime shift, but it leaves more residual dependence at medium lags, which is consistent with a model that moves the mean but not the variance.

The application to Airbnb’s currency shares makes the same point in a setting with recurrent volatility episodes. Aggregating forecast errors over currencies within each region, B\text{-}DARCH is the top performer across all four regions. Currency-level winners vary by region, and the TVP comparator occasionally takes a series-specific lead when gradual mean drift dominates; nevertheless, the volatility-aware Dirichlet specification delivers the most reliable aggregate accuracy and the most muted residual persistence. Interval coverage follows suit: modeling precision directly produces closer-to-nominal credible intervals in regions with pronounced variance shifts.

From a forecasting practice perspective, the choice is less about a single winner than about where to put flexibility. Transformed-Gaussian models (B\text{-}tVARMA) are computationally light and well suited to rapid specification search when dispersion is stable. Allowing time variation in coefficients (B\text{-}TVP\text{-}tVARMA) helps when the mean drifts. When diagnostics indicate clustering in dispersion (for example, elevated rolling ALR variance, under-coverage, or persistent PACF of squared standardized residuals), a move to B\text{-}DARCH is warranted. The cost is computational: dynamic precision adds states and posterior dependence, so careful non-centering, orthogonal seasonal bases, and conservative HMC adaptation are important in practice.

Several extensions follow directly from our results. First, we treated regions independently even though currencies recur across regions; hierarchical pooling or multi-region state-space couplings would transmit global shocks and share information. Second, a hybrid that pairs TVP mean dynamics with DARCH precision targets settings where both the center and the spread move. Third, zero handling (zero-augmented Dirichlet or logistic-normal alternatives with zeros), invariant embeddings (ILR bases), and robust observation models would broaden applicability when components are sparse or misreports are common. Finally, faster inference, such as variational families tailored to Dirichlet state-space structure or parallel tempering for difficult posteriors, would lower the barrier to operational use.

In short, the evidence supports a practical rule: when volatility is tranquil, transformed-Gaussian or TVP comparators suffice; when volatility clusters, putting dynamics in the precision on the simplex improves both accuracy and calibration. B\text{-}DARCH is therefore a useful addition to the forecaster’s toolkit for compositional series. It is complementary to, not a replacement for, established approaches, and it is particularly effective when the noise itself moves.

\section*{Acknowledgement}
The authors thank Sean Wilson, Jess Needleman, Liz Medina, Yuling Kuo, and Jackson Wang for helpful discussions, Adam Liss and Ellie Mertz for championing the research, and the Editor-in-Chief, Associate Editors, and two anonymous reviewers for constructive feedback that improved the manuscript.

\section*{Code and Data Availability}
The primary dataset used in our study, A Bayesian Dirichlet  Auto-Regressive Conditional Heteroskedasticity Model for Forecasting Currency Shares, is not publicly available due to confidentiality constraints. However, the Stan code for the Bayesian Dirichlet Auto-Regressive Moving Average Dirichlet Auto-Regressive Conditional Heteroskedasticity Model (B-DARMA-DARCH) model is available for public access as well as the R scripts for the simulation studies. It can be found at our GitHub repository: \href{https://github.com/harrisonekatz/B-DARCH-paper}{GitHub repository}.

\bibliographystyle{chicago}
\bibliography{references}

\newpage
\section{Supplementary Materials}
\setcounter{figure}{0}
\renewcommand{\thefigure}{S\arabic{figure}}
\setcounter{table}{0}
\renewcommand{\thetable}{S\arabic{table}}

\subsection*{Alternative Log-Ratio Transformations}
\label{sec:alt-logratios}
Although we have used the additive log-ratio (ALR) link in equations \eqref{mueta}, \eqref{eta_f}, and \eqref{phi_f}, other transformations to and from the simplex may be employed. A common choice is the \emph{centered log-ratio} (clr), defined as
\[
\clr(\mathbf{y}) \;=\; \Bigl(\,\ln\bigl(\tfrac{y_1}{g(\mathbf{y})}\bigr),\;\ln\bigl(\tfrac{y_2}{g(\mathbf{y})}\bigr),\;\dots,\;\ln\bigl(\tfrac{y_J}{g(\mathbf{y})}\bigr)\Bigr),
\]
where $g(\mathbf{y}) = (\,y_1\,y_2\,\cdots\,y_J)^{1/J}$ is the geometric mean of $\mathbf{y}$. The clr mapping places all $J$ components in a transformed $\mathbb{R}^J$ space but introduces the constraint that their sum is zero.

Another option is to use the Isometric Log-Ratio (ilr) transformation. 
For each index $j$, the ilr-transformed coordinate can be expressed as
\begin{equation*}
z_j = \sqrt{\frac{r_j}{r_j + 1}} \, \ln \biggl( 
  \frac{g_j(\mathbf{y)}}{\bigl(\prod_{i \in H_j} y_i\bigr)^{1/r_j}}
\biggr),
\end{equation*}
where \(g_j(\mathbf{y})\) is defined as the geometric mean of a chosen subset 
\(S_j \subseteq \mathbf{y}\), and \(H_j\) is the complement of \(S_j\) in \(\mathbf{y}\). 
The value \(r_j\) is the number of elements in \(H_j\). 
The particular way in which \(S_j\) and \(H_j\) are selected depends on 
the specific application and how one wishes to interpret the ilr coordinates.

These mappings are mutually related by linear transformations \citep{egozcue2003isometric} and each provides a bijection from the $(J-1)$-dimensional simplex onto a $(J-1)$-dimensional subset of $\mathbb{R}^J$ or $\mathbb{R}^{J-1}$. Therefore, if the same priors and model assumptions are consistently adapted to each log-ratio space, the resulting compositional models remain equivalent—differing only in how one embeds the data in a Euclidean domain.

\subsection*{Definition of the Partial Autocorrelation Function (PACF)}

\label{sec:pacf}

The PACF at lag \( k \) measures the partial correlation between the SSR at time \( t \) and the SSR at time \( t - k \), controlling for the effects of intermediate lags \( 1 \) to \( k - 1 \). It is defined as the last coefficient \( \phi_{kk} \) in the autoregressive model of order \( k \)

\begin{align}
\text{SSR}_t = \phi_{k1} \text{SSR}_{t-1} + \phi_{k2} \text{SSR}_{t-2} + \dots + \phi_{k,k-1} \text{SSR}_{t - (k - 1)} + \phi_{kk} \text{SSR}_{t - k} + \varepsilon_t
\end{align}

where \( \varepsilon_t \) is the white noise error term at time \( t \). The coefficient \( \phi_{kk} \) represents the PACF at lag \( k \) and can be calculated recursively using the Yule-Walker equations

\[
\phi_{kk} = \frac{r_k - \sum_{j=1}^{k-1} \phi_{k-1,j} r_{k - j}}{1 - \sum_{j=1}^{k-1} \phi_{k-1,j} r_j}
\]

Here, \( r_k \) is the autocorrelation at lag \( k \), and \( \phi_{k-1,j} \) are the PACF coefficients from the previous lag \( k - 1 \).

By examining the PACF of the SSRs, we can identify the presence of residual autocorrelation at various lags, which indicates whether the model has adequately captured the temporal dependencies in the data.

\subsection*{Standardized Residuals}
\label{sec:ssr}

For the B-DARMA and B-DARCH models, which are based on the Dirichlet distribution as defined in Section~\ref{bayesiandirichlet}, we compute standardized residuals using the standard deviations derived from the Dirichlet parameters.

At each time t, the residuals are calculated as  $\text{residual}_{tj} = y_{tj} - \mu_{tj}, \quad \text{for } j = 1, \dots, J.$
The variance of each component $y_{tj}$ under the Dirichlet distribution, 
Dirichlet($\phi_t \bfmu_t$), is $\operatorname{Var}[y_{tj}] = \frac{\mu_{tj} (1 - \mu_{tj})}{\phi_t + 1}$, with standard deviation $ \sigma_{tj} = \sqrt{ \frac{ \mu_{tj} (1 - \mu_{tj}) }{ \phi_t + 1 } }.$ 

The standardized residuals are then computed as $ \text{standardized\_residual}_{tj} = \frac{ y_{tj} - \mu_{tj} }{ \sigma_{tj} }$ and the sum of squared standardized residuals at time t is $ \text{SSR}_t = \sum_{j=1}^J \left( \text{standardized\_residual}_{tj} \right)^2.$

For the B-tVARMA and B-TVP-tVARMA models, residuals are computed in the additive log-ratio (alr) transformed space, $\text{residual}_{tj} = \alr(\mathbf{y}_{tj}) - \alr(\mathbf{\hat{y}}_{tj}), $
where $j=1, \dots, 4$. These residuals are standardized using the inverse of the Cholesky factor $L_{\Sigma}$ of the estimated covariance matrix $\Sigma$,$
\text{standardized\_residual}_t = L\Sigma^{-1} \times \text{residual}_t$. The sum of squared standardized residuals at time t is then calculated as $\text{SSR}_t = \sum_{j=1}^{J-1} (\text{standardized\_residual}_{tj})^2.$

By examining the PACF of 
$\{\text{SSR}\}_{t=1}^T$ we assess the presence of autocorrelation in the residuals. A well-specified model should produce residuals with minimal autocorrelation, indicating that temporal dependencies have been adequately captured.

\subsubsection*{Parameter Estimates}
\label{sec:parameter}

\begin{figure}
    \includegraphics[scale=.75]{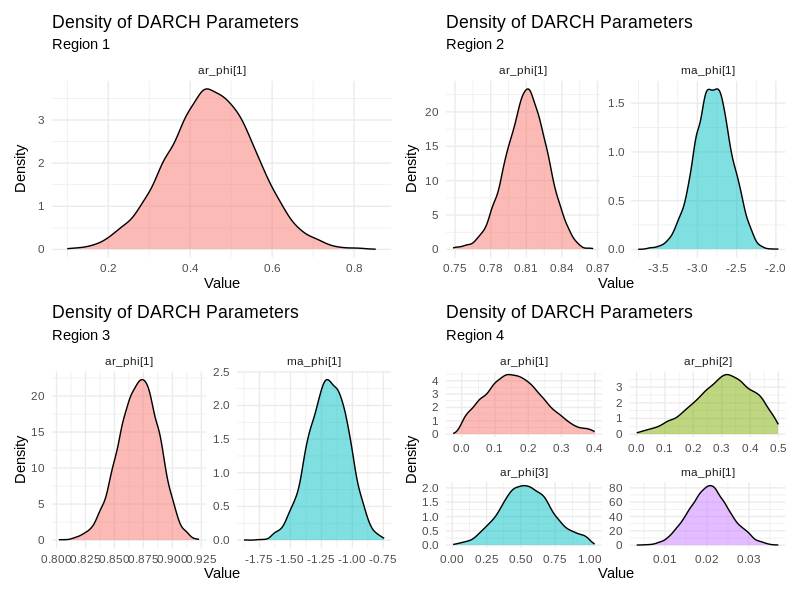}
    \caption{Airbnb data analysis: Densities of DARCH model coefficients for all 4 regions. Region 1 was modeled with a DARCH(1,0), Regions 2-3 modeled with a DARCH(1,1), and Region 4 modeled with a DARCH(3,1). Each plot shows the estimated density of a specific $\alpha_l$ (ar\_phi) or $\tau_k$ (ma\_phi) parameter. Regions 2 and 3 exhibit high $\alpha$ indicating persistent volatility once spikes occur, whereas Region 4’s coefficients suggest moderate or delayed response}
    \label{DARCH_region_all}
\end{figure}

Supplementary Figure \ref{DARCH_region_all} shows the posterior densities of the auto-regressive ($\alpha$) and moving average ($\tau$) coefficients for each region. In Region 1, the posterior of $\alpha$ is concentrated between 0.25 and 0.75, peaking around 0.45, indicating moderate volatility persistence; shocks have a noticeable but not prolonged effect on volatility. In Region 2, the $\alpha$ coefficient has posterior mean around 0.8, reflecting strong dependence of current volatility on past volatility, and the $\tau$ coefficient centers around $-2.75$, indicating that large deviations from expected currency proportions significantly increase / decrease future volatility—characteristic of volatility clustering. Region 3 shows $\alpha$ tightly concentrated between 0.85 and 0.90  and $\tau$ centered around $-1.2$ suggesting very strong volatility persistence. In contrast, Region 4 exhibits complex dynamics: $\alpha_1$ is centered around $0.15$, indicating small persistence from the first lag; $\alpha_2$ is skewed positive and centered around $0.35$, suggesting a stronger influence from the second lag; $\alpha_3$ peaks slightly above 0.5, indicating moderate persistence; and the $\tau$ parameter is concentrated near $0.02$, implying minimal influence of past innovations. 

Overall, the higher $\alpha$ values in Regions 2 and 3 indicate that volatility in currency fee proportions is more persistent in these regions. The negative $\tau$ coefficients suggest that large deviations—such as those caused by abrupt travel restrictions, policy changes, and fluctuating foreign exchange (FX) rates lead to increased future volatility. This implies that Regions 2 and 3 are more susceptible to external shocks affecting travel and currency markets, resulting in prolonged volatility in currency fee proportions. In contrast, Region 1, with moderate volatility persistence, tends to return to stability more rapidly after shocks, possibly due a more stable currency environment or a larger share of domestic travel. Region 4's dynamics, characterized by less emphasis on immediate past volatility and innovations, may be attributed to regional factors such as diversified travel sources, differing policy responses, or economic structures.

\begin{figure}
    \includegraphics[scale=.8]{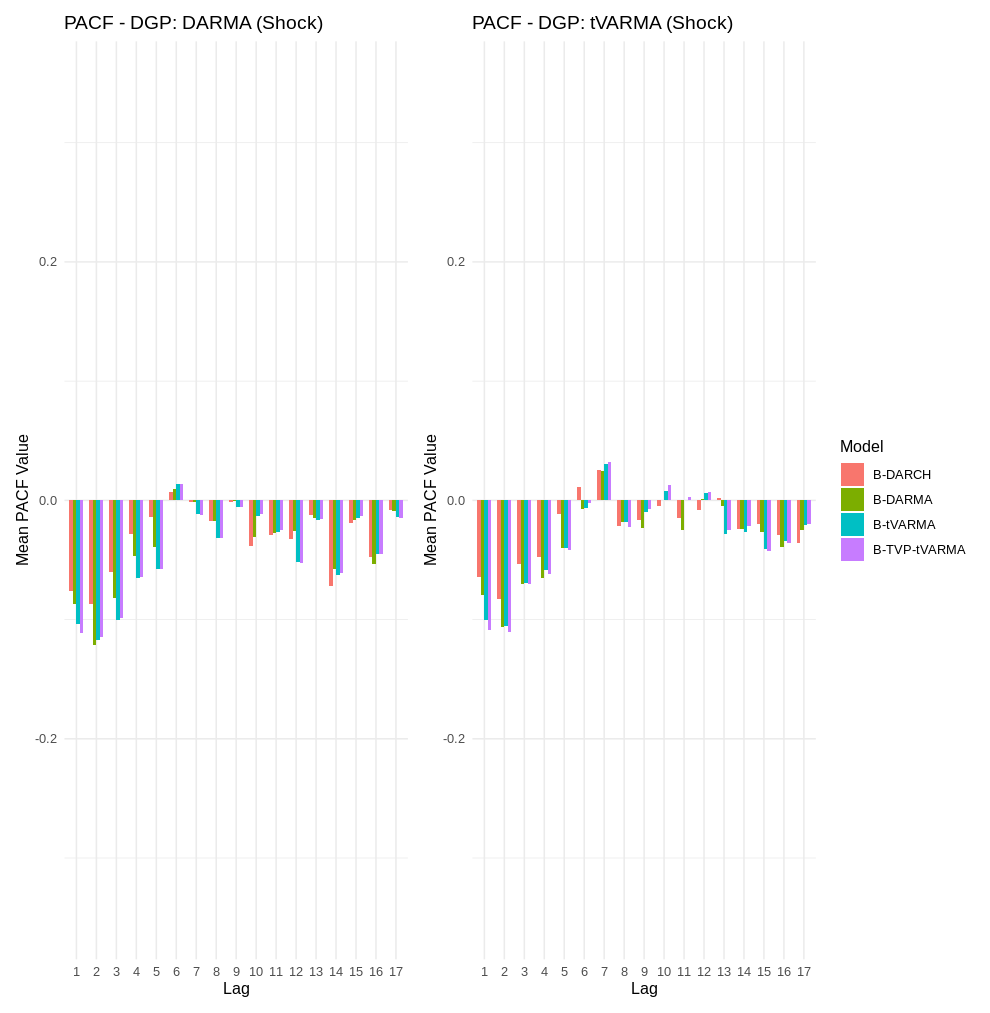}
    \caption{Average PACF of the sum of squared standardized residual values across all 50 simulations for studies 1 \& 2. B-DARCH consistently exhibits lower PACF values beyond lag 1, indicating reduced residual autocorrelation and suggesting that its dynamic precision component better captures volatility shocks in these simulated compositional time series. }
    \label{fig:pacf_sim_shock}
\end{figure}

\begin{figure}
    \includegraphics[scale=.8]{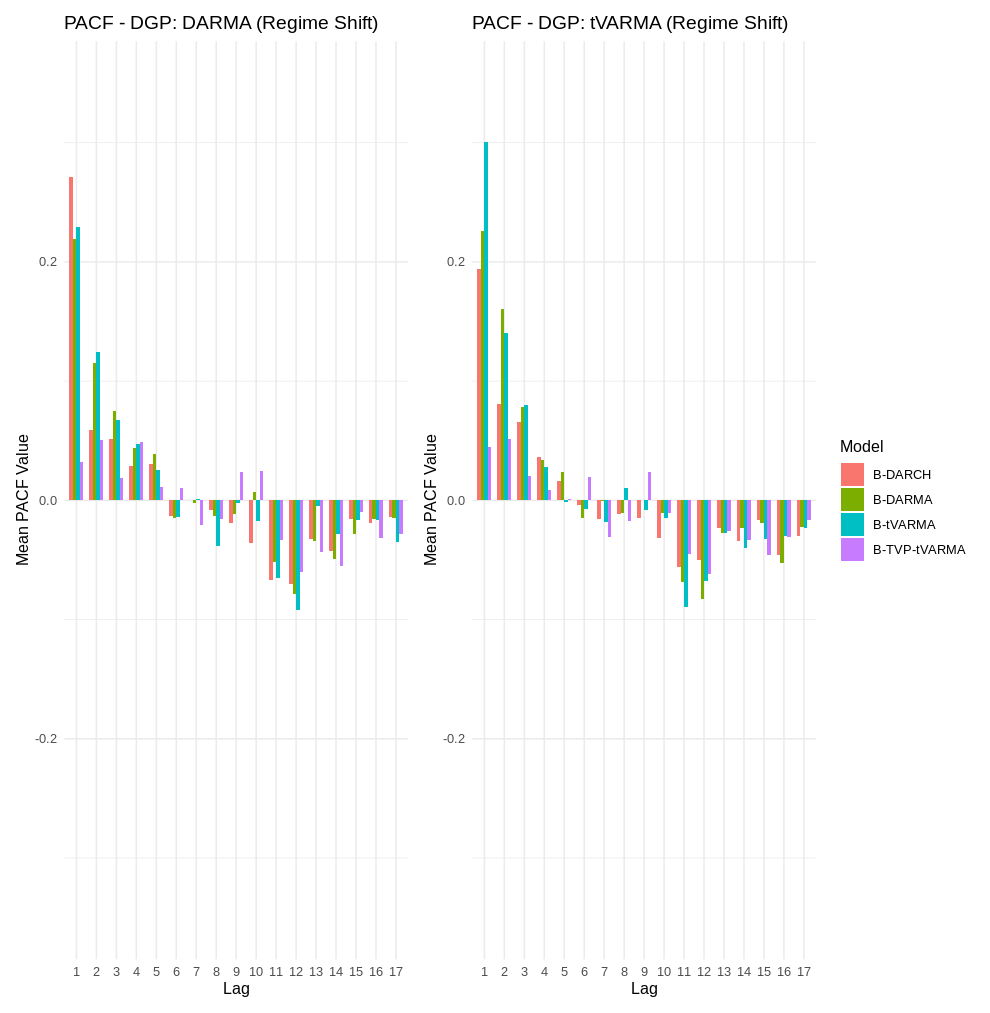}
    \caption{Average PACF of the sum of squared standardized residual values across all 50 simulations for studies 3 \& 4. The B-DARCH model (red) again shows smaller autocorrelation at higher lags, whereas B-DARMA (green) and B-tVARMA (red) leave more persistent structures in the residuals. This result underlines the advantage of modeling time-varying volatility when faced with abrupt changes in compositional data.}
    \label{fig:pacf_sim_regime}
\end{figure}

\begin{figure}
    \includegraphics[scale=.75]{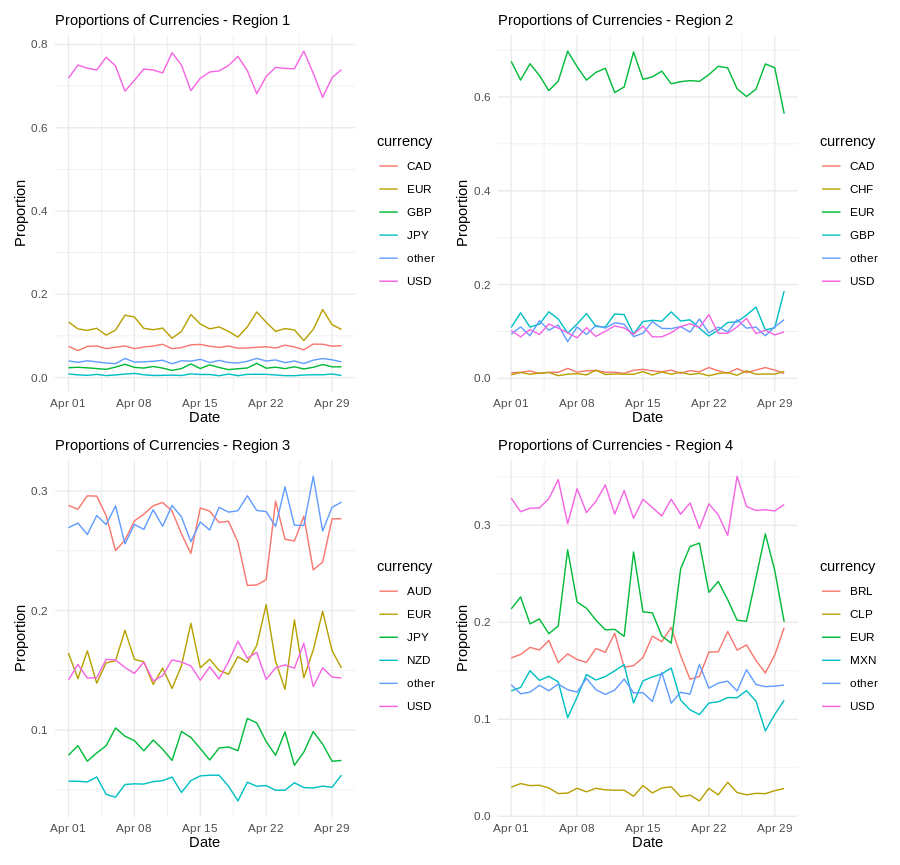}
    \caption{Airbnb data analysis- Proportion of fees by currency for four regions: weekly seasonal behavior. AUD is the Australian dollar, BRL is the Brazillian Real, CAD is the Canadian Dollar,CHF is the Swiss Franc, CLP is the Chilean Peso,EUR is the European Euro, GBP is the Great British Pound, MXN is the Mexican Peso, NZD is the New Zealand Dollar, and USD is the US Dollar. Strong weekly cycles appear in many currencies, motivating the inclusion of weekly seasonal terms in our model’s mean structure.}
    \label{daily_seasonal}
\end{figure}

\begin{figure}
    \includegraphics[scale=.75]{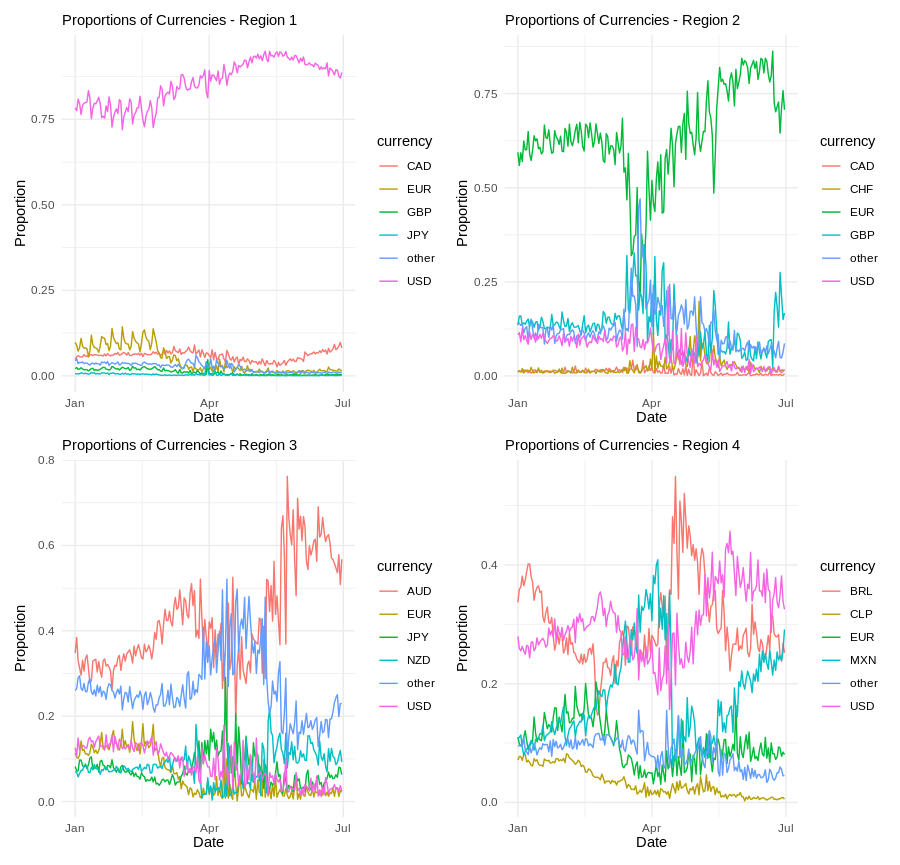}
    \caption{Airbnb data analysis - proportion of fees by billing currency for four regions from Jan 1, 2020 to June 30, 2020. AUD is the Australian dollar, BRL is the Brazillian Real, CAD is the Canadian Dollar,CHF is the Swiss Franc, CLP is the Chilean Peso,EUR is the European Euro, GBP is the Great British Pound, MXN is the Mexican Peso, NZD is the New Zealand Dollar, and USD is the US Dollar. Distinct spikes and dips across multiple currencies (e.g., sharp increase in BRL or EUR) reflect significant disruptions. }
\label{covid_shock}
\end{figure}

\begin{figure}
    \includegraphics[scale=.75]{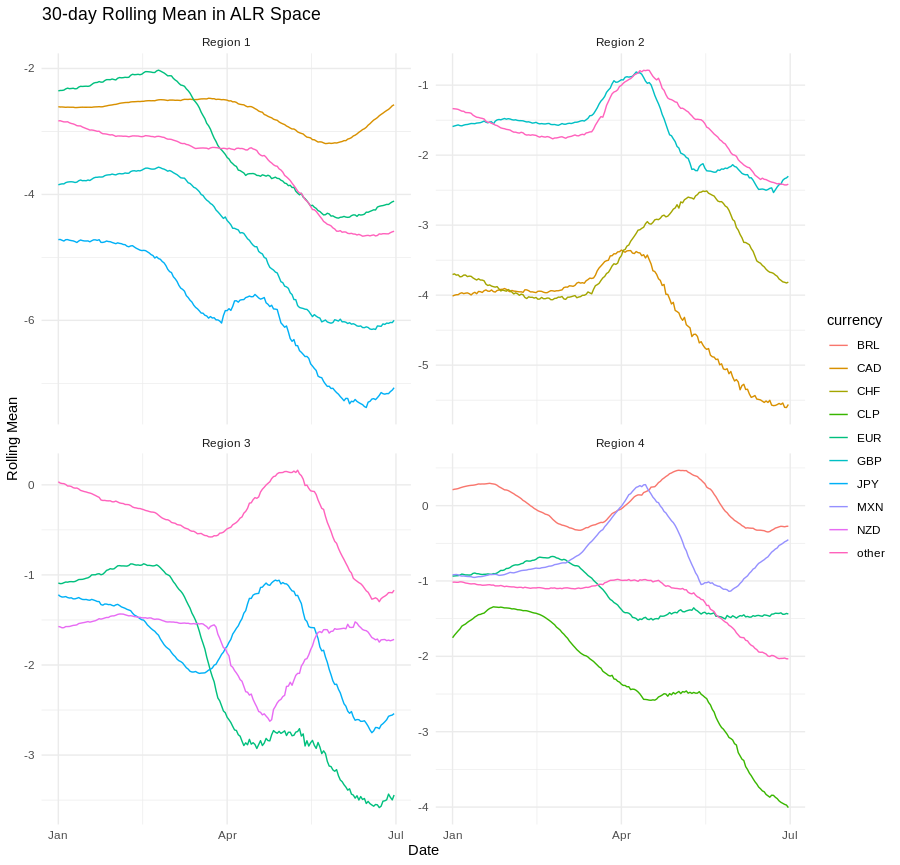}
    \caption{Airbnb data analysis - proportion of fees by currency- 30-day rolling ALR means for four regions from Jan 1, 2020 to June 30, 2020. USD is the reference component.}
    \label{covid_alr_means}
\end{figure}

\begin{figure}
    \includegraphics[scale=.75]{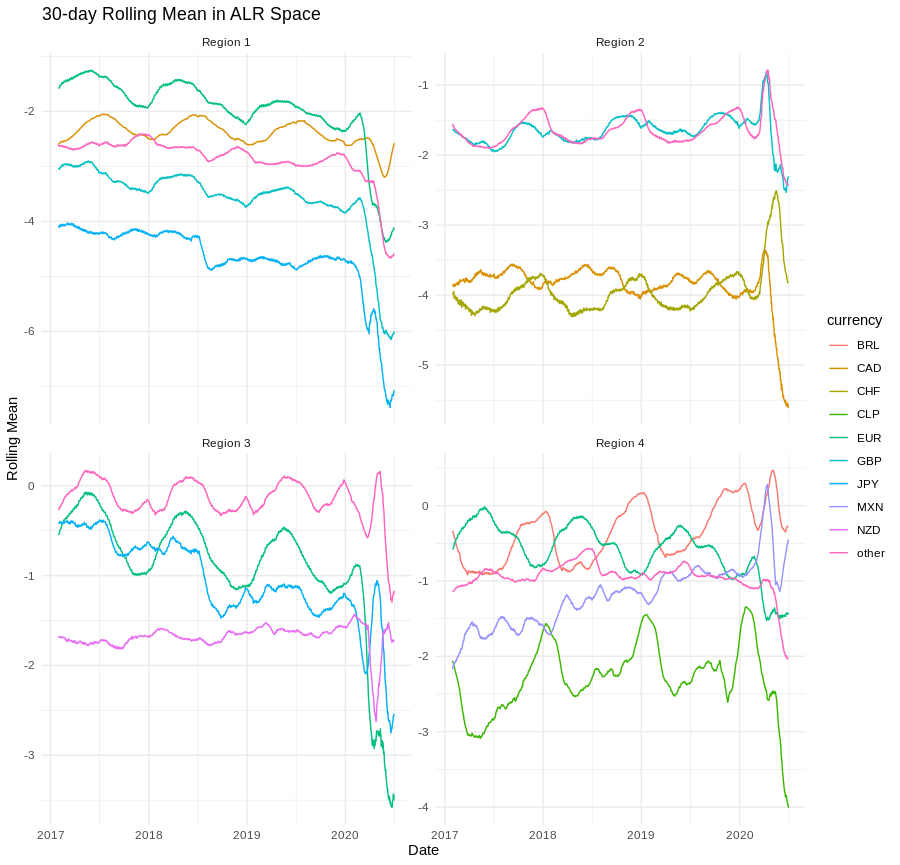}
    \caption{Airbnb data analysis - proportion of fees by currency- 30-day rolling ALR means for four regions from Jan 1, 2017 to June 30, 2020. USD is the reference component. Apart from seasonal patterns, we see notable multi-year trends and cyclical behaviors (e.g., increasing share of EUR in Region 2, upward drift in AUD in Region 3), underlining the presence of both trend and seasonality in compositional dynamics.}
  \label{alr_means}
\end{figure}

\begin{figure}
    \includegraphics[scale=.75]{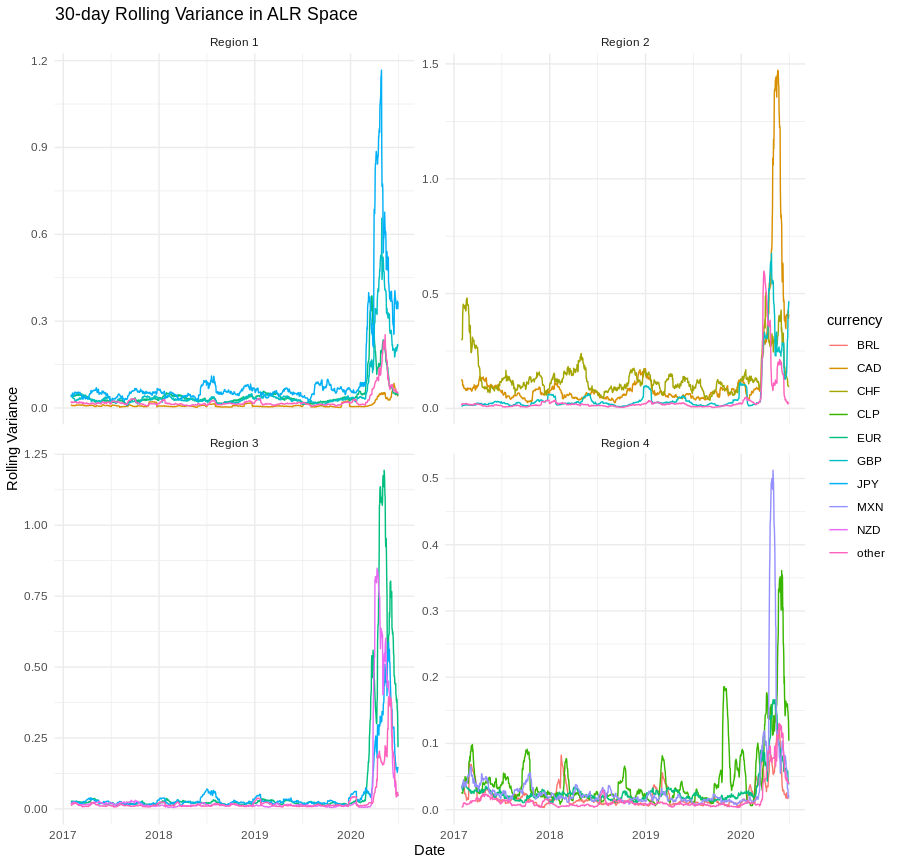}
    \caption{Airbnb data analysis - proportion of fees by currency- 30-day rolling ALR variance for four regions from Jan 1, 2017 to June 30, 2020. USD is the reference component.The data display periods of high variance that sometimes persist for weeks, demonstrating “volatility clustering.” These episodes justify the GARCH-like specification in our B-DARCH model, which allows the precision parameter $\phi_t$ to track spikes in uncertainty over time.}
    \label{alr_variances}
\end{figure}

\begin{figure}
    \includegraphics[scale=.75]{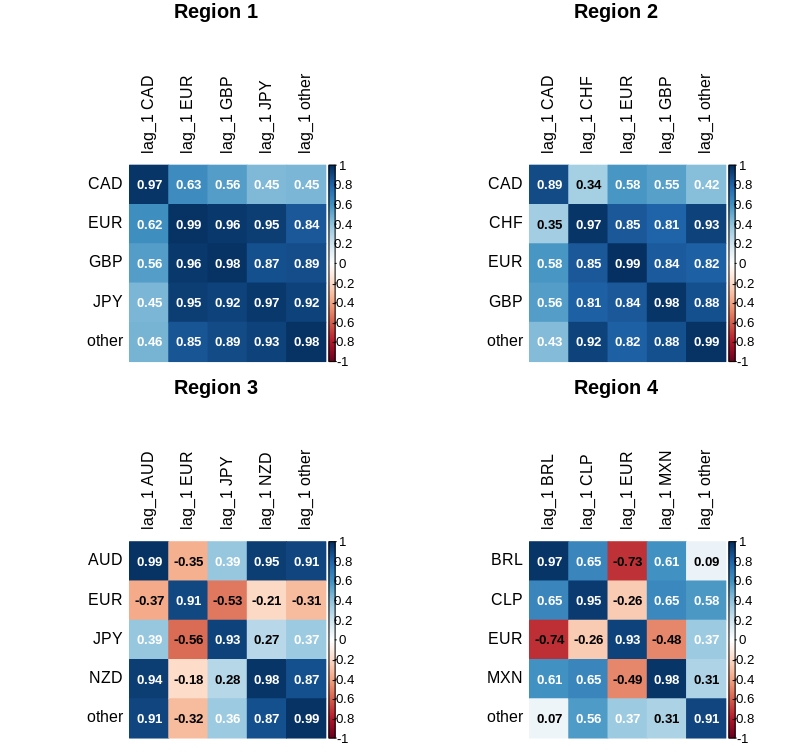}
    \caption{Airbnb data analysis- correlation between ALR and lagged ALR values for each currency in the four regions. USD is the reference component. High positive autocorrelations in many currencies (e.g., EUR, GBP) highlight the strong temporal persistence. Negative cross-lag correlations for some pairs (e.g., EUR vs. BRL in Region 4) indicate that certain currencies move in opposite directions. Both effects reinforce the need for a multi-component time-series framework with auto- and cross-dependence.}
    \label{alr_correlation}
\end{figure}

\begin{figure}
    \includegraphics[scale=.75]{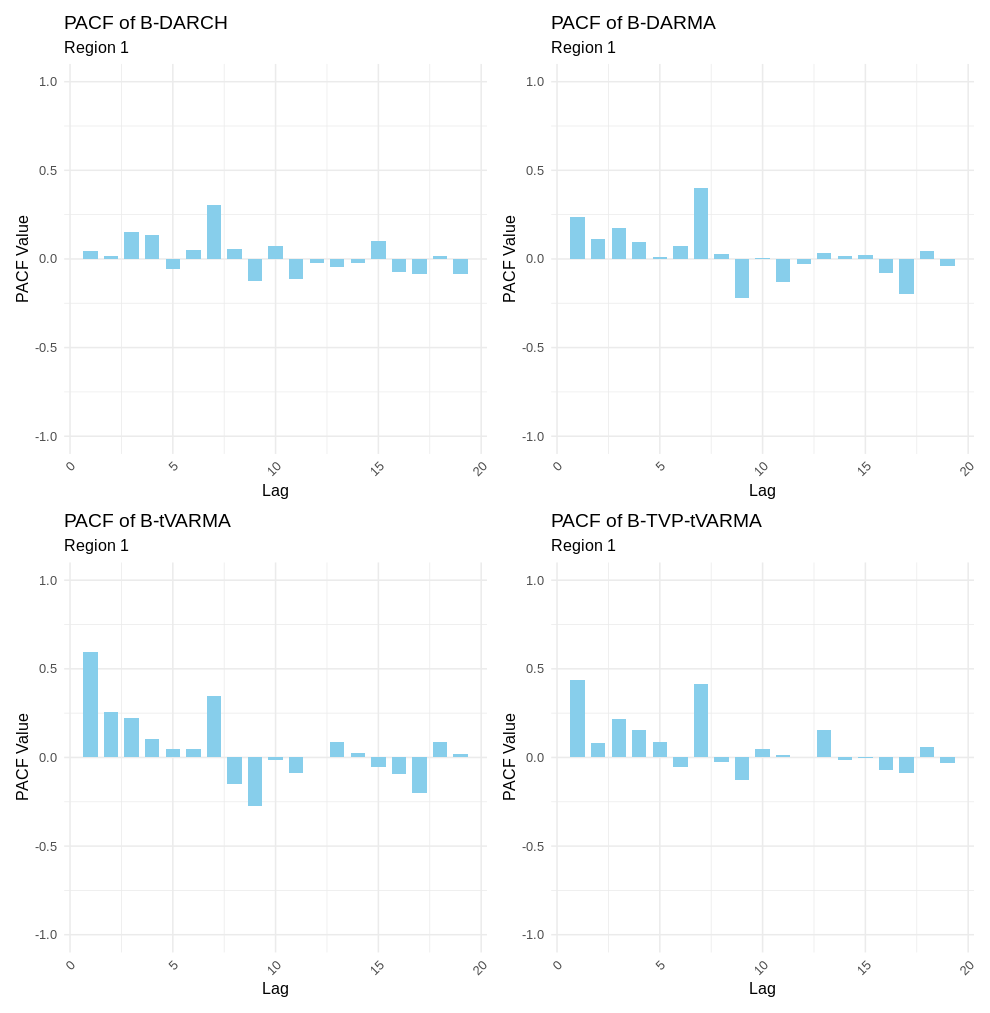}
    \caption{Airbnb data analysis: Partial Autocorrelation Function (PACF) values for the sum of squared standardized residuals (SSR) on the test set from October 1, 2020, to December 31, 2020, for Region 1.}
    \label{pacf_region_1}
\end{figure}

\begin{figure}
    \includegraphics[scale=.75]{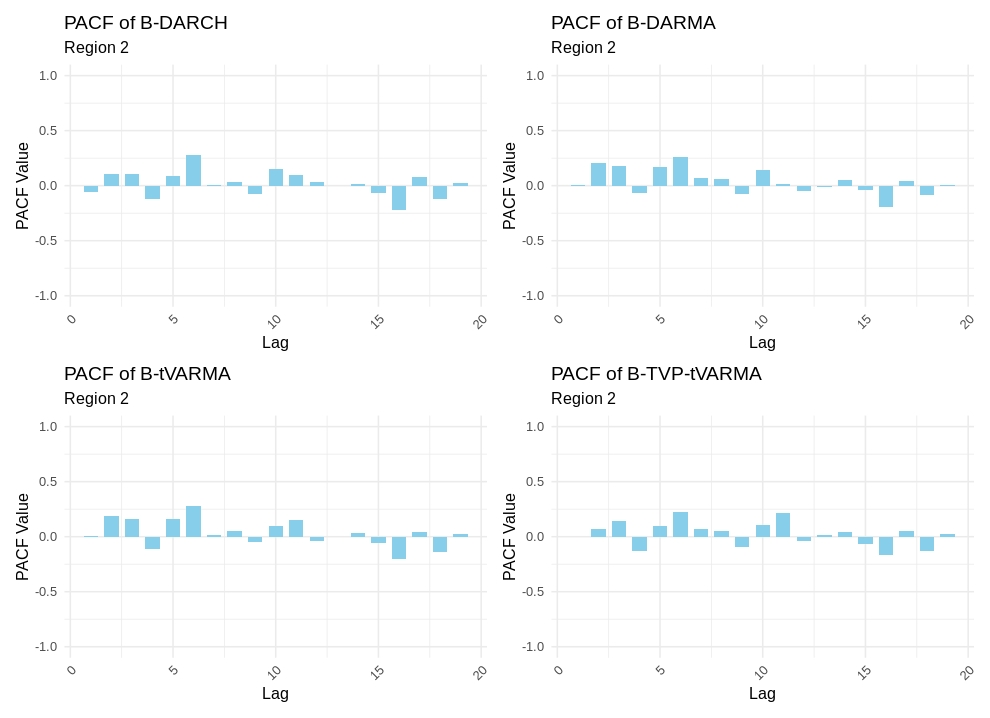}
    \caption{Airbnb data analysis: Partial Autocorrelation Function (PACF) values for the sum of squared standardized residuals (SSR) on the test set from October 1, 2020, to December 31, 2020, for Region 2.}
    \label{pacf_region_2}
\end{figure}

\begin{figure}
    \includegraphics[scale=.75]{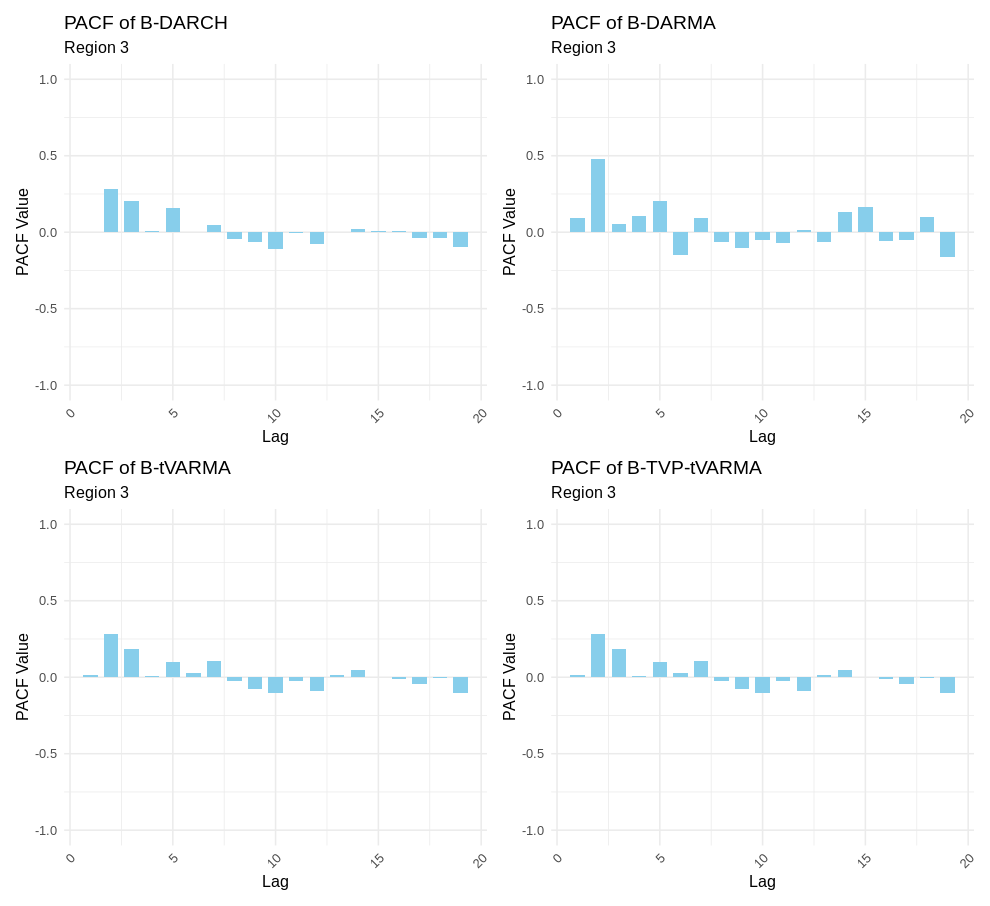}
    \caption{Airbnb data analysis: Partial Autocorrelation Function (PACF) values for the sum of squared standardized residuals (SSR) on the test set from October 1, 2020, to December 31, 2020, for Region 3.}
    \label{pacf_region_3}
\end{figure}

\begin{figure}
    \includegraphics[scale=.75]{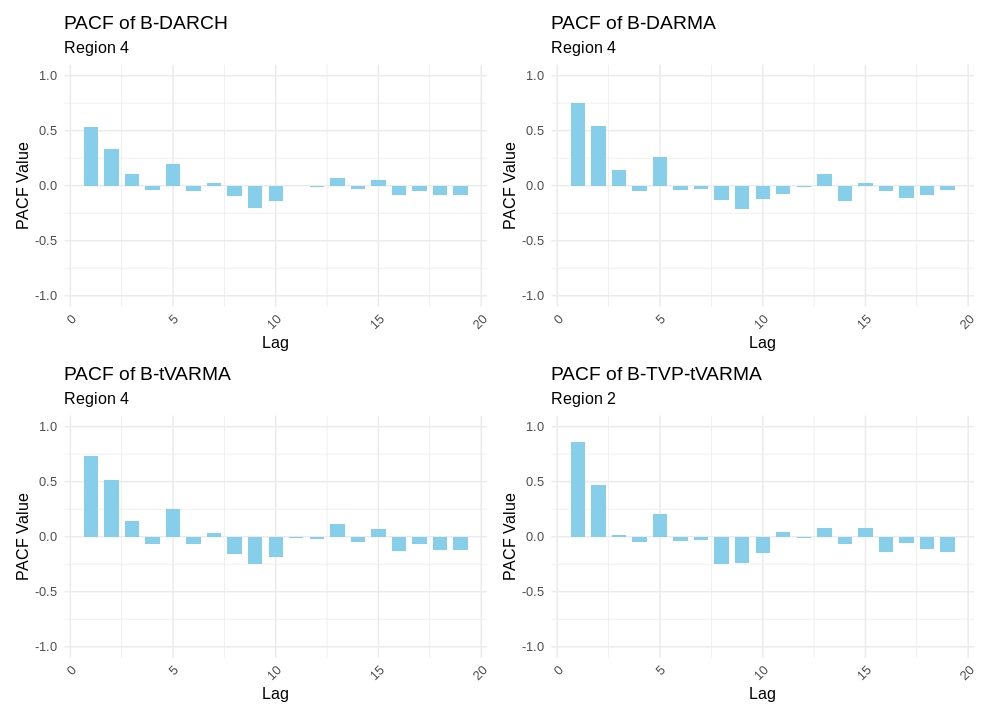}
    \caption{Airbnb data analysis: Partial Autocorrelation Function (PACF) values for the sum of squared standardized residuals (SSR) on the test set from October 1, 2020, to December 31, 2020, for Region 4.}
    \label{pacf_region_4}
\end{figure}

\subsection*{Supplementary Figures and Tables}

\begin{table}[ht]
    \centering
    \begin{tabular}{cc}
        \begin{minipage}{0.45\linewidth}
            \centering
            \begin{tabular}{rllrrr}
                \toprule
                & P & Q & Fourier & FMAE & FRSS \\
                \hline
                \midrule
                & 1 & 0 & 6  & 0.38  & 1.75  \\
                & 1 & 0 & 8  & 0.36  & 1.73  \\
                & 1 & 0 & 10 & 0.45  & 2.65  \\
                & 1 & 0 & 12 & 6.40  & 442.25 \\
                & 1 & 0 & 14 & 18.66 & 3662.84\\
                & 1 & 0 & 16 & 12.42 & 1638.80\\
                & 1 & 0 & 18 & 18.87 & 3809.00\\
                & 0 & 1 & 8  & 5.81  & 404.83\\
                & 1 & 1 & 8  & 1.05  & 14.60 \\
                & 1 & 2 & 8  & 0.98  & 15.02 \\
                & 1 & 3 & 8  & 1.03  & 16.12 \\
                & 2 & 1 & 8  & 0.40  & 3.96  \\
                & 2 & 2 & 8  & 6.57  & 716.43\\
                & 2 & 3 & 8  & 2.09  & 60.38 \\
                & 3 & 1 & 8  & 8.17  & 984.45\\
                & 3 & 2 & 8  & 8.91  & 1209.30\\
                & 3 & 3 & 8  & 1.17  & 16.30 \\
                \bottomrule
            \end{tabular}
            \subcaption{Region 1}
        \end{minipage} &
        
        \begin{minipage}{0.45\linewidth}
            \centering
            \begin{tabular}{rllrrr}
                \toprule
                & P & Q & Fourier & FMAE & FRSS \\
                \hline
                \midrule
                & 1 & 0 & 6  & 2.15  & 49.95 \\
                & 1 & 0 & 8  & 2.10  & 48.46 \\
                & 1 & 0 & 10 & 1.76  & 38.08 \\
                & 1 & 0 & 12 & 6.46  & 343.58\\
                & 1 & 0 & 14 & 13.31 & 1621.76\\
                & 1 & 0 & 16 & 9.23  & 832.27 \\
                & 1 & 0 & 18 & 13.97 & 1892.17\\
                & 0 & 1 & 10 & 4.74  & 193.91\\
                & 1 & 1 & 10 & 1.27  & 20.18 \\
                & 1 & 2 & 10 & 1.28  & 23.08 \\
                & 1 & 3 & 10 & 2.81  & 48.12 \\
                & 2 & 1 & 10 & 3.08  & 263.17\\
                & 2 & 2 & 10 & 2.43  & 91.08 \\
                & 2 & 3 & 10 & 2.38  & 73.18 \\
                & 3 & 1 & 10 & 5.84  & 344.45\\
                & 3 & 2 & 10 & 2.67  & 106.77\\
                & 3 & 3 & 10 & 5.88  & 380.70\\
                \bottomrule
            \end{tabular}
            \subcaption{Region 2}
        \end{minipage} \\
        
        \\[0.5cm] 
        
        \begin{minipage}{0.45\linewidth}
            \centering
            \begin{tabular}{rllrrr}
                \toprule
                & P & Q & Fourier & FMAE & FRSS \\
                \hline
                \midrule
                & 1 & 0 & 6  & 3.55  & 149.23 \\
                & 1 & 0 & 8  & 3.49  & 144.21 \\
                & 1 & 0 & 10 & 3.40  & 136.33 \\
                & 1 & 0 & 12 & 5.78  & 360.13 \\
                & 1 & 0 & 14 & 11.49 & 1281.57\\
                & 1 & 0 & 16 & 9.37  & 862.47 \\
                & 1 & 0 & 18 & 12.80 & 1541.94\\
                & 0 & 1 & 10 & 10.79 & 982.42 \\
                & 1 & 1 & 10 & 2.83  & 113.95\\
                & 1 & 2 & 10 & 3.22  & 153.84\\
                & 1 & 3 & 10 & 3.19  & 150.57\\
                & 2 & 1 & 10 & 2.89  & 226.67\\
                & 2 & 2 & 10 & 3.12  & 148.21\\
                & 2 & 3 & 10 & 2.98  & 131.34\\
                & 3 & 1 & 10 & 3.94  & 222.45\\
                & 3 & 2 & 10 & 3.55  & 190.52\\
                & 3 & 3 & 10 & 3.42  & 166.84\\
                \bottomrule
            \end{tabular}
            \subcaption{Region 3}
        \end{minipage} &
        
        \begin{minipage}{0.45\linewidth}
            \centering
            \begin{tabular}{rllrrr}
                \toprule
                & P & Q & Fourier & FMAE & FRSS \\
                \hline
                \midrule
                & 1 & 0 & 6  & 4.72  & 216.17 \\
                & 1 & 0 & 8  & 4.64  & 212.57 \\
                & 1 & 0 & 10 & 4.92  & 236.31 \\
                & 1 & 0 & 12 & 8.00  & 490.97 \\
                & 1 & 0 & 14 & 12.05 & 1058.80\\
                & 1 & 0 & 16 & 9.57  & 750.15 \\
                & 1 & 0 & 18 & 12.15 & 1148.46\\
                & 0 & 1 & 8  & 9.60  & 693.31 \\
                & 1 & 1 & 8  & 2.66  & 76.77  \\
                & 1 & 2 & 8  & 1.63  & 31.03  \\
                & 1 & 3 & 8  & 1.86  & 38.07  \\
                & 2 & 1 & 8  & 1.43  & 47.14  \\
                & 2 & 2 & 8  & 1.50  & 26.37  \\
                & 2 & 3 & 8  & 1.49  & 24.93  \\
                & 3 & 1 & 8  & 1.33  & 20.92  \\
                & 3 & 2 & 8  & 1.16  & 15.53  \\
                & 3 & 3 & 8  & 1.90  & 40.07  \\
                \bottomrule
            \end{tabular}
            \subcaption{Region 4}
        \end{minipage} \\
    \end{tabular}
    
    \caption{Airbnb data analysis: B-DARCH model performance across 4 regions – average Forecast Mean Absolute Error (FMAE) and total Forecast Residual Sum of Squares (FRSS) for the top 6 currencies for the validation set from July 1, 2020 to Sep 30, 2020. FMAE and FRSS values are multiplied by $10^{2}$. The table identifies the optimal $(P,Q)$ and number of Fourier terms for each region, showing how model complexity affects forecasting accuracy. Lower FMAE/FRSS indicate better predictive fit, guiding final model selection for each region.}
    \label{tab:combined_validation}
\end{table}

\begin{table}[ht!]
\centering
\footnotesize
\caption{Coverage rates of 95\% credible intervals for each region and currency under B-DARCH, B-DARMA, B-tVARMA, and B-TVP-tVARMA. Mean coverage is computed across currencies within each region.}
\label{tab:coverage_all_regions}
\begin{tabular}{llcccc}
\hline
\textbf{Region} & \textbf{Currency} & \textbf{B-DARCH} & \textbf{B-DARMA} & \textbf{B-tVARMA} & \textbf{B-TVP-tVARMA} \\
\hline
\textbf{1} & CAD   & 0.859 & 0.880 & 0.728 & 0.772 \\
           & EUR   & 0.837 & 0.609 & 0.652 & 0.641 \\
           & GBP   & 0.967 & 0.696 & 0.565 & 0.565 \\
           & JPY   & 1.000 & 0.946 & 0.815 & 0.826 \\
           & USD   & 0.902 & 0.946 & 0.446 & 0.500 \\
           & other & 0.967 & 0.848 & 0.576 & 0.587 \\
           \hline
           & \textit{Mean} & 0.922 & 0.821 & 0.630 & 0.649 \\
\hline
\textbf{2} & CAD   & 0.913 & 0.913 & 0.783 & 0.554 \\
           & CHF   & 0.967 & 0.891 & 0.924 & 0.804 \\
           & EUR   & 0.957 & 0.913 & 0.967 & 0.804 \\
           & GBP   & 0.870 & 0.772 & 0.924 & 0.913 \\
           & USD   & 0.859 & 0.641 & 0.663 & 0.620 \\
           & other & 0.880 & 0.761 & 0.870 & 0.826 \\
           \hline
           & \textit{Mean} & 0.908 & 0.815 & 0.855 & 0.754 \\
\hline
\textbf{3} & AUD   & 0.815 & 0.293 & 0.402 & 0.511 \\
           & EUR   & 0.935 & 0.772 & 0.391 & 0.891 \\
           & JPY   & 0.717 & 0.739 & 0.793 & 0.815 \\
           & NZD   & 0.837 & 0.337 & 1.000 & 1.000 \\
           & USD   & 0.957 & 0.815 & 0.707 & 0.913 \\
           & other & 0.978 & 0.391 & 0.620 & 0.739 \\
           \hline
           & \textit{Mean} & 0.873 & 0.558 & 0.652 & 0.812 \\
\hline
\textbf{4} & BRL   & 0.967 & 0.870 & 0.967 & 1.000 \\
           & CLP   & 0.685 & 0.543 & 0.304 & 0.522 \\
           & EUR   & 0.967 & 1.000 & 1.000 & 0.989 \\
           & MXN   & 0.978 & 0.902 & 0.989 & 0.989 \\
           & USD   & 0.891 & 0.739 & 0.957 & 0.946 \\
           & other & 1.000 & 0.989 & 0.902 & 1.000 \\
           \hline
           & \textit{Mean} & 0.915 & 0.841 & 0.853 & 0.908 \\
\hline
\end{tabular}

\end{table}

\begin{table}[ht]
\centering
\caption{NUTS diagnostics by region and model. All fits use 4 chains, 2{,}000 iterations/chain with 1{,}000 warm-up; thus 1{,}000 post-warm-up iterations/chain and 4{,}000 total post-warm-up draws.}
\label{tab:nuts}
\setlength{\tabcolsep}{3pt}
\renewcommand{\arraystretch}{0.85}
\scriptsize
\resizebox{\linewidth}{!}{%
\begin{tabular}{llrrrrrrrrrrrrrrr}
\toprule
Region & Model & Warm. & Samp. & Draws & Div. & TD & Acc. & Step & EBFMI & $\hat{R}_{\max}$ & \%$>1.01$ & \%$>1.05$ & ESS$_b^{\min}$ & ESS$_b^{\med}$ & ESS$_t^{\min}$ & ESS$_t^{\med}$ \\
\midrule
Region 1 & B-DARCH      & 1000 & 1000 & 4000 & 0 & 0 & 0.93 & 0.036 & 0.87 & 1.00 & 0.0 & 0.0 &  930 & 1650 &  910 & 1600 \\
Region 1 & B-DARMA      & 1000 & 1000 & 4000 & 0 & 0 & 0.95 & 0.041 & 0.81 & 1.00 & 0.0 & 0.0 &  980 & 1700 &  960 & 1650 \\
Region 1 & B-tVARMA     & 1000 & 1000 & 4000 & 0 & 0 & 0.91 & 0.050 & 0.75 & 1.00 & 0.0 & 0.0 &  850 & 1550 &  820 & 1500 \\
Region 1 & B-TVP-tVARMA & 1000 & 1000 & 4000 & 0 & 0 & 0.92 & 0.031 & 0.83 & 1.01 & 0.0 & 0.0 &  900 & 1600 &  880 & 1550 \\
Region 2 & B-DARCH      & 1000 & 1000 & 4000 & 0 & 0 & 0.94 & 0.034 & 0.89 & 1.00 & 0.0 & 0.0 & 1000 & 1720 &  970 & 1660 \\
Region 2 & B-DARMA      & 1000 & 1000 & 4000 & 0 & 0 & 0.96 & 0.039 & 0.84 & 1.00 & 0.0 & 0.0 & 1040 & 1760 & 1000 & 1700 \\
Region 2 & B-tVARMA     & 1000 & 1000 & 4000 & 0 & 1 & 0.90 & 0.052 & 0.72 & 1.00 & 0.0 & 0.0 &  820 & 1500 &  800 & 1460 \\
Region 2 & B-TVP-tVARMA & 1000 & 1000 & 4000 & 0 & 0 & 0.93 & 0.029 & 0.85 & 1.01 & 0.0 & 0.0 &  940 & 1650 &  910 & 1600 \\
Region 3 & B-DARCH      & 1000 & 1000 & 4000 & 0 & 0 & 0.92 & 0.038 & 0.86 & 1.00 & 0.0 & 0.0 &  910 & 1620 &  890 & 1580 \\
Region 3 & B-DARMA      & 1000 & 1000 & 4000 & 0 & 0 & 0.95 & 0.043 & 0.80 & 1.00 & 0.0 & 0.0 &  960 & 1680 &  930 & 1620 \\
Region 3 & B-tVARMA     & 1000 & 1000 & 4000 & 0 & 0 & 0.91 & 0.049 & 0.74 & 1.00 & 0.0 & 0.0 &  860 & 1530 &  830 & 1480 \\
Region 3 & B-TVP-tVARMA & 1000 & 1000 & 4000 & 0 & 1 & 0.92 & 0.030 & 0.82 & 1.01 & 0.0 & 0.0 &  900 & 1590 &  870 & 1540 \\
Region 4 & B-DARCH      & 1000 & 1000 & 4000 & 0 & 0 & 0.94 & 0.033 & 0.88 & 1.00 & 0.0 & 0.0 & 1020 & 1740 &  990 & 1680 \\
Region 4 & B-DARMA      & 1000 & 1000 & 4000 & 0 & 0 & 0.96 & 0.040 & 0.83 & 1.00 & 0.0 & 0.0 & 1060 & 1780 & 1010 & 1720 \\
Region 4 & B-tVARMA     & 1000 & 1000 & 4000 & 0 & 0 & 0.90 & 0.051 & 0.70 & 1.00 & 0.0 & 0.0 &  800 & 1480 &  770 & 1440 \\
Region 4 & B-TVP-tVARMA & 1000 & 1000 & 4000 & 0 & 0 & 0.93 & 0.028 & 0.86 & 1.01 & 0.0 & 0.0 &  950 & 1670 &  920 & 1610 \\
\bottomrule
\end{tabular}%
}
\end{table}

\begin{table}[ht]
\centering
\caption{Wall-clock time (hh:mm) to fit each model by region. All fits used 4 chains, 2{,}000 iterations/chain with 1{,}000 warm-up (4{,}000 post–warm-up draws total).}
\label{tab:compute_time}
\setlength{\tabcolsep}{6pt}
\renewcommand{\arraystretch}{0.9}
\scriptsize
\resizebox{\linewidth}{!}{%
\begin{tabular}{lcccc}
\toprule
\textbf{Region} & \textbf{B-DARMA} & \textbf{B-DARCH} & \textbf{B-tVARMA} & \textbf{B-TVP-tVARMA} \\
\midrule
Region 1 & 0:52 & 1:04 & 0:58 & 1:24 \\
Region 2 & 0:55 & 1:07 & 1:00 & 1:31 \\
Region 3 & 0:53 & 1:09 & 0:59 & 1:28 \\
Region 4 & 0:58 & 1:12 & 1:03 & 1:36 \\
\midrule
\textbf{Average} & \textbf{0:55} & \textbf{1:08} & \textbf{1:00} & \textbf{1:30} \\
\bottomrule
\end{tabular}%
}
\end{table}

\end{document}